\documentclass{aa}
\usepackage{graphicx}

\usepackage{txfonts}
\usepackage{booktabs}
\usepackage{latexsym}
\usepackage{array}

\usepackage{amsfonts,amssymb,amsmath}
\usepackage{braket}

\usepackage[colorlinks=true, allcolors=blue]{hyperref}

\newcommand{\new}[1]{{#1}}
\newcommand{\neww}[1]{{#1}}

\begin{document}


\title{
\neww{Radio afterglows of binary neutron star mergers: a population study for \neww{current and future gravitational wave observing runs}}\\
}

\author{R. Duque, F. Daigne \& R. Mochkovitch}

\authorrunning{R. Duque, et al.}
\titlerunning{Predictions for radio afterglows of binary neutron star mergers}

\institute{Sorbonne Universit\'e, CNRS, UMR 7095, Institut d'Astrophysique de Paris, 98 bis boulevard Arago, 75014 Paris, France}

\date{Received: 21 May 2019 / Accepted: 30 August 2019}

\abstract
{Following the historical observations of GW170817 and its multi-wavelength afterglow, more radio afterglows from neutron star mergers are expected in the future as counterparts to gravitational wave inspiral signals.
We wish to describe these events using our current knowledge of the population of neutron star mergers coming from gamma-ray burst science, and taking into account the sensitivities of current and future gravitational wave and radio detectors.
\neww{We combine analytical models for the merger gravitational wave and radio afterglow signals to a population model prescribing the energetics, circum-merger density and other relevant parameters of the mergers.}
We report the expected distributions of observables (distance, orientation, afterglow peak time/flux, etc.) from future events and study how these can be used to further probe the population of binary neutron stars, their mergers and related outflows during future observing campaigns. In the case of the O3 run of the LIGO-Virgo Collaboration, the radio afterglow of one third of gravitational-wave-detected mergers should be detectable (and detected if the source is localized thanks to the kilonova counterpart) by the Very Large Array, and these events should have viewing angles similar to that of GW170817. These findings confirm the radio afterglow as a powerful insight on these events, though some key afterglow-related techniques, such as very long baseline interferometry imaging of the merger remnant, may no longer be feasible as the gravitational wave horizon increases.}


\maketitle

\section{Introduction}
The first detection of the gravitational waves (GW) from the inspiral phase of a binary neutron star merger \citep{23, 76} was followed by all three electromagnetic counterparts expected after the coalescence: a short gamma-ray burst (GRB) \citep{52, 137}, its multi-wavelength afterglow (AG) in the X-ray \citep{133, 134, 135, 13}, radio \citep{12} and optical
bands \citep{8} and an optical-IR thermal transient source
\citep{122, 126, 121},
which allowed to localize the event with sub-arcsecond precision in the S0-type galaxy NGC4993 at a distance of $40.7\pm 3.3$~Mpc \citep{33, 22}.
This thermal transient showed evidence for heating from $r$-process nucleosynthesis \citep{127, 128, 131, 53}, classifying it as a ``kilonova'' (KN), the first with such detailed observations.

According to estimates on joint GW+GRB events by \cite{67}, such a combined detection of GW+GRB+AG+KN should remain rare.
Indeed, GRB170817A would not have been detected at a distance larger than 50~Mpc \citep{52} or from a viewing angle larger than~$\sim~25^\circ$. 

\new{In the context of the present O3 and future observing runs of the LIGO-Virgo Collaboration (LVC), we would like to know what to expect regarding the afterglows of future binary neutron star (BNS) mergers: the rates of GW and joint GW+AG events, the distributions of event observables such as distance $D$, viewing angle $\theta_{\rm v}$, radio afterglow time of peak and peak flux $t_p$ and $F_p$ and \new{remnant} proper motion $\mu$, and the sensitivity of these to the population's characteristics and to the multi-messenger detector configuration.}


In this paper
, we \new{approach} these questions from the angle of a population study. Precisely, we start from prior knowledge on binary neutron star mergers from short GRB science and observations of the 170817 event. This allows us to prescribe some intrinsic variability in mergers of the population such as their kinetic energy, external medium density, shock microphysical conditions, etc. We then calculate the \new{radio} afterglows arising from this population of mergers in the \new{hypothesis} of a jet-dominated afterglow emission, and finally apply \new{thresholds} of current and future GW and \new{radio} detectors to determine within the population those events which will be \new{detectable} jointly in \new{the} GW and \new{radio} domains. \new{By describing and studying this jointly-detectable population,} \new{we make predictions for future observing runs}.


Such a population model for short GRB afterglows was already considered by \cite{166, 167}, with detailed numerical models for the multi-band afterglow and GW signals \new{and using uninformative (e.g. uniform or log-uniform) event parameter distributions.} Moreover, the discussion \new{therein} was centered on the parameter-space constraints one can derive from detections or non-detections of the afterglow in various electromagnetic bands, and on the conditions \new{necessary} to observe afterglows from off-axis lines-of-sight. Here, we will rather discuss the effect of the population parameters and detector configurations on the expected population of afterglows.

More recently, this approach was also followed by \cite{145}, and again our study differs in that it accounts for some prior knowledge on short GRBs (such as their luminosity function) and observations from GW170817 to sharpen our predictions on the events to come. This allows a detailed study of the impact of the population parameters on the predicted observations.

Our study relies on the following assumptions:
\begin{itemize}
    \item \textit{In most cases the merger produces a successful central jet}. This is clearly a strong assumption that will have to be validated by future observations. Recent studies based on the population of short GRBs \citep{67} and the dynamics of jets interacting with merger ejecta \citep{111} suggest that successful jets could be common in these phenomena. 
    \item \new{\textit{Around its peak, the afterglow is dominated by the contribution of the jet's core}}. In this case, the afterglow peak flux depends on the jet's \new{core's} parameters (\new{core} isotropic kinetic energy $E_\mathrm{iso,c}$ and opening angle $\theta_{\rm j}$, shock microphysics parameters $\epsilon_{\rm e}$, $\epsilon_{\rm B}$ and $p$), on its environment (external medium density $n$), and \new{on} its viewing conditions (the distance and viewing angle). The \new{population's} distributions of most of these quantities remain quite uncertain. \new{Our knowledge of these} relies on the afterglow fitting of a limited sample of short GRB afterglows with known distance \citep[e.g.][]{28, 118}, on the expectation that short GRBs generally occur in a low density environment and on the results obtained from the (up to now) single \new{multi-messenger} event GRB170817A \cite[e.g.][]{88}.    
\end{itemize}

This publication is organized as follows. In Sec.~\ref{sec:2} we describe our \new{methods} to calculate the afterglow emission and detail the input parameters of our population model. In Sec.~\ref{criteria} we give the criteria for GW and radio afterglow detection we apply to \new{deduce} the jointly-detectable population. In Sec.~\ref{sec:results} we report general results on this population, describe its \new{observable} characteristics and study their sensitivity to the detector configurations and to the population model parameters. In Sec.~\ref{sec:discussion}, we discuss our results in the context of future multi-messenger campaigns and how they may help to interpret observations therein.

\section{A population model for binary neutron star merger afterglows}
\label{sec:2}

\subsection{Radio afterglows from structured jets}

The multi-wavelength afterglow of GW170817 is associated with the deceleration of a
structured relativistic jet
emitted by the central source formed \new{in} the merger, and more precisely to the synchrotron emission of electrons accelerated by the forward shock propagating in the external medium. 
It was observed for more than 300 days. \new{Its} time evolution is similar at all wavelengths,
indicating that the emission 
\new{was} produced in the same spectral regime of the slow cooling synchrotron process, i.e. $\nu_\mathrm{m}< \nu
< \nu_\mathrm{c}$,
and that the non-thermal distribution of shock-accelerated electrons \new{assumed} a non-evolving slope 
$p\sim 2.2$ \citep{83, 80, 88}.
The observed slow rise of the afterglow is clear evidence of the lateral structure of the outflow, which is observed off-axis with a viewing angle $\theta_\mathrm{v}\sim 15-25^\circ$. \new{Initially}, the observed flux is dominated by the beamed emission from mildly-relativistic/mildly-energetic material coming towards the observer \citep{5}. \new{As the jet decelerates}, relativistic beaming becomes less efficient and 
regions closer to the jet axis start to contribute to the flux. The peak \new{of the afterglow} is reached when the jet's core, \new{which is open} to an angle $\theta_{\rm j}$ of a few degrees, is revealed. This occurs when the jet's core's Lorentz factor has \new{decreased} down to $\Gamma\sim 1/\theta_\mathrm{v}$.

\new{An} alternative possibility is a quasi-spherical mildly relativistic ejecta with a steep radial structure \citep{9, 
42, 149, 5, 
7
}. This is now strongly disfavored by radio very long baseline interferometry (VLBI) observations, \new{which showed} an apparent superluminal motion \new{of the remnant} with $\beta_\mathrm{app}\sim 4$ \citep{79} and \new{constrained} the angular size of the radio source \new{to being} below $2\, \mathrm{mas}$ \citep{110}, \new{thus confirming the emergence of a relativistic core jet at the peak of the afterglow}.

The lateral structure of the jet is constrained by observations: best fits are obtained for a sharp decay of the kinetic energy and Lorentz factor with the angle $\theta$ from the jet axis \citep{151, 154, 
150, 152, 17}. 
Typically, for a top-hat core jet 
with an opening angle $\theta_\mathrm{j}$, isotropic equivalent kinetic energy $E_\mathrm{iso,c}=4\pi \epsilon_\mathrm{c}$ and Lorentz factor $\Gamma_\mathrm{c}$ surrounded by a power-law-structured sheath with the following kinetic energy per solid angle $\epsilon(\theta)$ and Lorentz factor $\Gamma_0(\theta)$, 

\begin{equation}
\epsilon(\theta) = \epsilon_\mathrm{c} \, \left\lbrace\begin{array}{lc}
1 & \mathrm{if}\, \theta\le \theta_\mathrm{j}\\
\left(\theta/\theta_\mathrm{j}\right)^{-a} & \mathrm{if}\, \theta\ge \theta_\mathrm{j}\\
\end{array}\right.
\label{eq:structure1}
\end{equation}
and
\begin{equation}
\Gamma_0(\theta) = 1+\left(\Gamma_\mathrm{c}-1\right) \, \left\lbrace\begin{array}{lc}
1 & \mathrm{if}\, \theta\le \theta_\mathrm{j}\\
\left(\theta/\theta_\mathrm{j}\right)^{-b} & \mathrm{if}\, \theta\ge \theta_\mathrm{j}\\
\end{array}\right.\, ,
\label{eq:structure2}
\end{equation}
\new{observations} require steep slopes $a,b\gtrsim 2$
\citep{7,110}.

\begin{figure*}
\resizebox{\hsize}{!}{\includegraphics{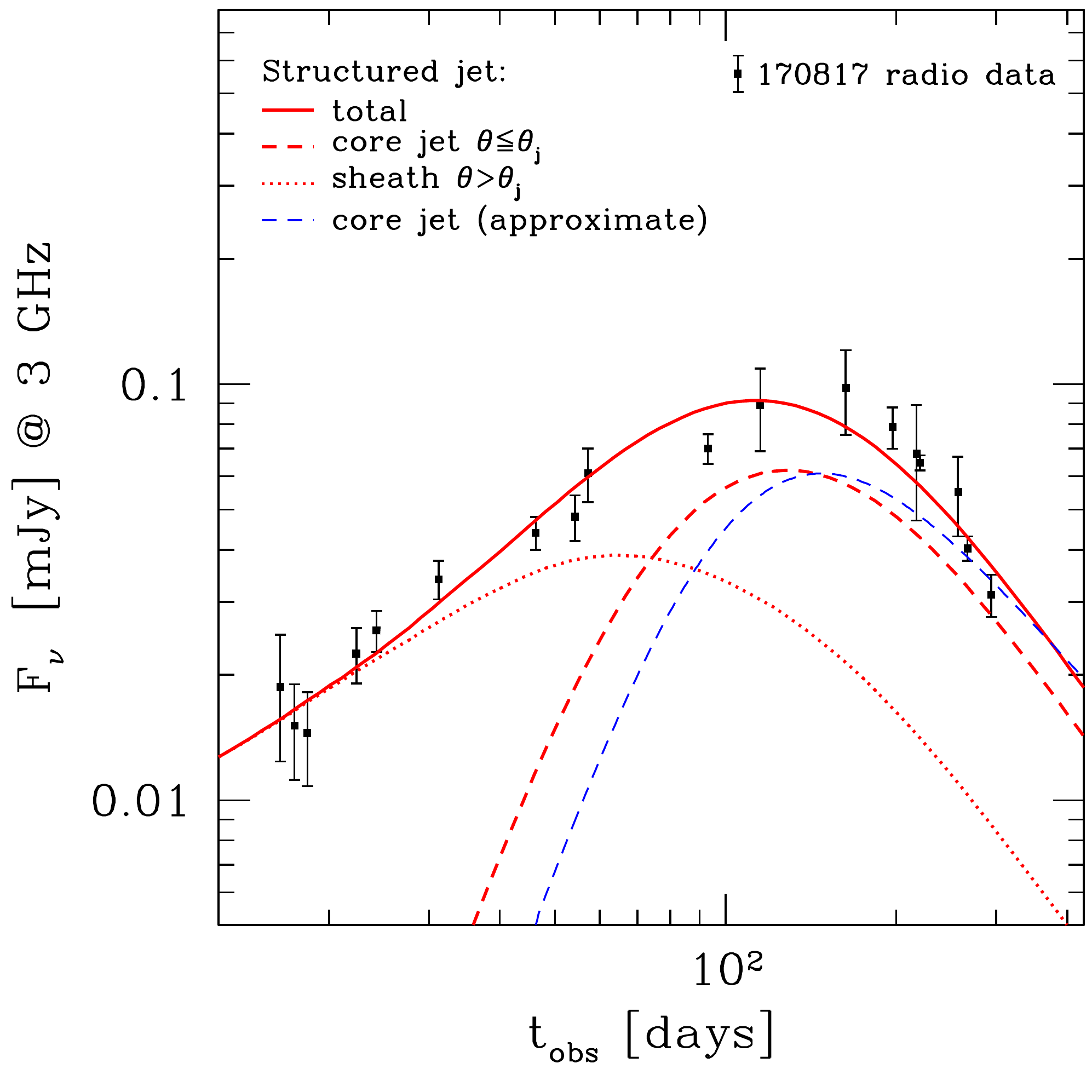}\includegraphics{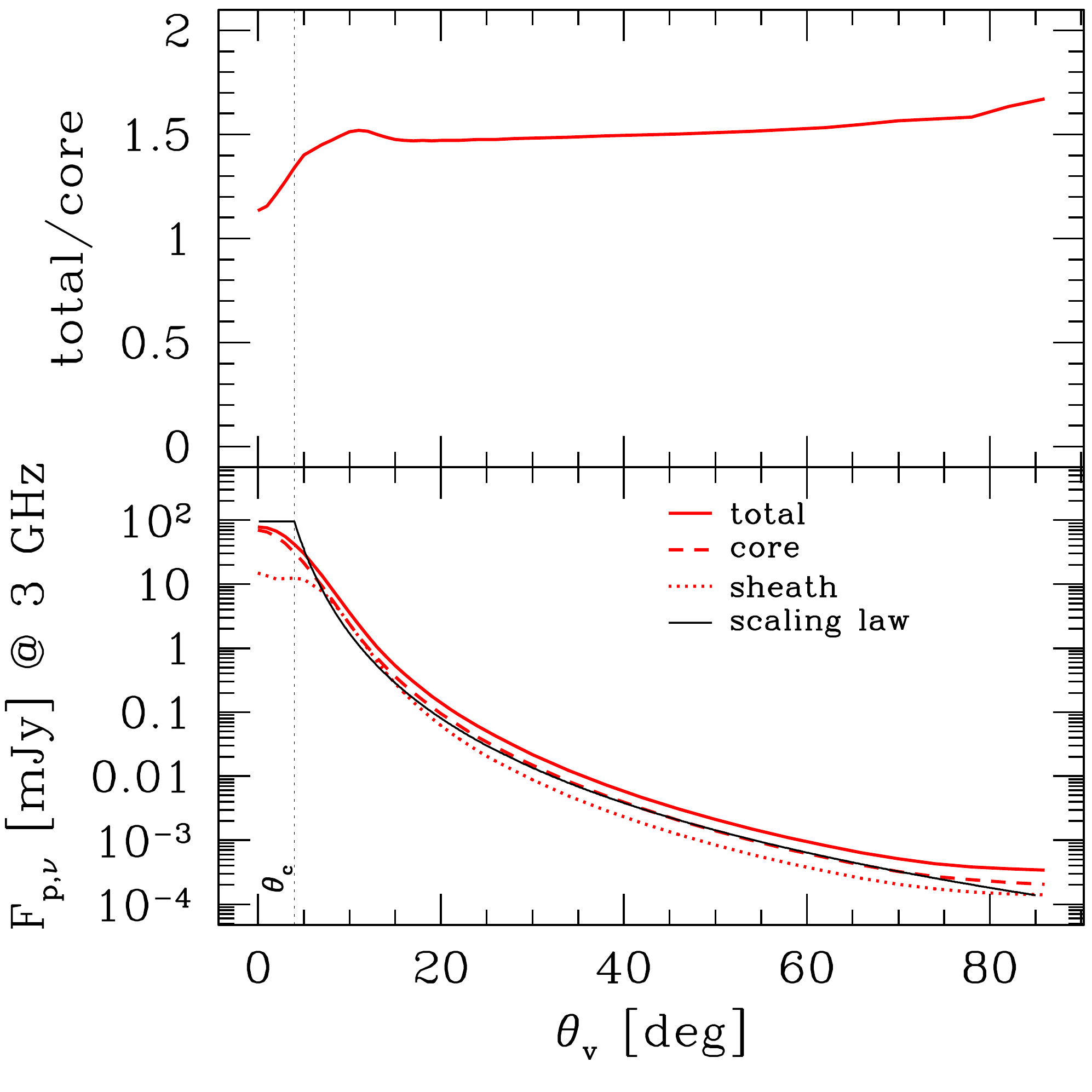}}
\caption{\textit{Left:} 3 GHz light curve of the afterglow from a structured jet at a
distance $D=42~\mathrm{Mpc}$, with a sharp power-law structure ($a=4.5$, $b=2.5$; see Eq.~\ref{eq:structure1}-\ref{eq:structure2}), a viewing angle $\theta_\mathrm{v}=22^\circ$, an external density $n=3\, 10^{-3}\, \mathrm{cm^{-3}}$, a core jet with $\theta_\mathrm{j}=4^\circ$, $\Gamma_\mathrm{c}=100$, $E_\mathrm{iso,c}=2\, 10^{52}\, \mathrm{erg}$, and with microphysics parameters $p=2.2$, $\epsilon_\mathrm{e}=0.1$ and $\epsilon_\mathrm{B}=10^{-4}$. The radio observations of GW170817 are also plotted and are compiled from \cite{12, 29, 17, 80, 5, 10}. The respective contributions of the core jet ($\theta\le\theta_\mathrm{j}$), the sheath and the total are plotted in dashed, dotted and solid red lines. The flux of the core jet computed with the simplified treatment described by Eq.~\ref{eq:FluxApp} is also plotted in dashed blue  line for comparison. \textit{Bottom right:} Peak flux  of the same structured jet as a function of the viewing angle $\theta_\mathrm{v}$ (solid red line), peak flux of the light curve from the core jet only (dashed red line) or from the sheath only (dotted red line), and peak flux of the core jet as computed using the scaling law in Eq.~\ref{eq:FP} (thin black line). \textit{Upper right:} Ratio of the peak flux from the 
whole outflow (core jet+sheath)
to that of the core jet only.}
\label{fig:radiofit}
\end{figure*}

An example of the radio lightcurves obtained for such a structured jet is plotted in Fig.~\ref{fig:radiofit} (left) using parameters typical of the various fits to the data that have been published \citep[see e.g.][]{7,88,110}. 
To \new{obtain} this lightcurve, the dynamics of the material at \new{different latitudes are computed independently}. At a latitude $\theta$, the deceleration radius \new{is}  $R_\mathrm{dec}(\theta) = \left(\frac{3\epsilon(\theta)}{\Gamma_0^2(\theta) n m_\mathrm{p} c^2} \right)^{1/3}$, which is constant in the core and slowly increases with $\theta$ outside the core ($R_\mathrm{dec}(\theta)\propto \theta^{(2b-a)/3}$). \new{For the values of $a$ and $b$ chosen in Fig.~\ref{fig:radiofit},  $\frac{2b-a}{3} = 0.17$.} 

Then, the synchrotron emission of shock-accelerated electrons in the shock comoving frame is assumed to follow the standard synchrotron slow-cooling spectrum  \citep{3} including self-absorption. Finally, the observed lightcurve is computed by summing the contributions of all latitudes on equal-arrival time surfaces, taking relativistic beaming and Doppler boosting into account.   

The separate contributions of the core jet and the sheath are plotted in Fig.~\ref{fig:radiofit} (left) and the emergence of the core at the peak is clearly visible. The evolution of the peak flux of the same structured jet is plotted in Fig.~\ref{fig:radiofit} (right) as a function of the viewing angle, as well as the ratio of the peak flux to the peak flux 
from the core jet only. Interestingly, this ratio is almost constant ($\sim 1.5$), \new{except for the on-axis} cases ($\theta_\mathrm{v}\le \theta_\mathrm{j}$) when the core is more dominant. 

As long as  the kinetic energy and the Lorentz factor decay steeply with $\theta$, the core jet is expected to dominate the flux at the peak whatever the viewing angle,  even if the precise value  of the total-to-core ratio at the peak may slightly vary depending on the details of the assumed lateral structure. 
Therefore, as our population model considers only the properties of the afterglow at the peak, i.e. when it can more easily be detected, it is enough in the following to compute only the contribution from the core jet, \new{keeping in mind} that it may slightly underestimate the total peak flux by a factor\footnote{\new{See Sec.~\ref{sec:discussion} for a short discussion of the effect of this on the fraction of detectable events.}} $\sim~1.5$. 

The contribution from a top-hat core jet can efficiently be computed using the approximation suggested by \citet{19} to avoid the full integration over equal-arrival time surfaces. Precisely, we compute the flux $F_\nu^\mathrm{iso}(t_\mathrm{obs})$ of a spherical ejecta with initial Lorentz factor $\Gamma_\mathrm{c}$ and kinetic energy $E_\mathrm{iso,c}$ as in \cite{1}, and correct it by defining the on-axis jet-breaked afterglow flux:
\begin{equation}
F^\mathrm{on}_\nu(t_\mathrm{obs}) = F_\nu^\mathrm{iso}(t_\mathrm{obs})\times\left\{\begin{array}{ll}
1 & \mathrm{if}\, \Gamma\theta_\mathrm{j}\le 1\\
\left( \Gamma\theta_\mathrm{j} \right)^{-2} & \mathrm{if}\, \Gamma\theta_\mathrm{j}\ge 1
\end{array}\right.
\end{equation}
and calculating the afterglow from any viewing angle as:
\begin{equation}
F_\nu(t_\mathrm{obs}) = \left\{\begin{array}{ll} F^\mathrm{on}_\nu(t_\mathrm{obs}) & \rm{if}~\theta_v < \theta_j~\rm{(on~axis)} \\ a^3 F^\mathrm{on}_{\nu/a}(b\, t_\mathrm{obs}) & \rm{if}~\theta_v > \theta_j~\rm{(off~axis)}\end{array}\right.
\label{eq:FluxApp}
\end{equation}
with $a=\frac{1-\beta}{1-\beta\cos{\left(\theta_\mathrm{v}-\theta_\mathrm{j}\right)}}$ 
and $b=\frac{1-\frac{R}{c t}}{1-\frac{R}{c t}\cos{\left(\theta_\mathrm{v}-\theta_\mathrm{j}\right)}}$ 
where $t$ is the source frame time. The latter correction, corresponding to 
 the ratio of on-axis/off-axis arrival times
is not included in \cite{19} and is found to improve the quality of the approximation.
The corresponding light curve is plotted in Fig.~\ref{fig:radiofit} (left, blue dashed line) and agrees well with the exact calculation.

\subsection{An analytical model for the afterglow peak properties}

\new{This afterglow calculation is standard for top-hat jets}. Therefore analytical expressions for the \new{afterglow} properties at the peak are available. They depend slightly on the assumptions for a possible late jet lateral expansion. If lateral expansion is taken into account like in standard GRB afterglow theory, the peak flux of the radio light curve scales as \citep{114}:
\begin{equation}
F_{p,\nu}  \propto\, E_{\rm iso,c} \,\theta_{\rm j}^2\,n^{\frac{p+1}{4}}\,\epsilon_{\rm e}^{p-1}\,\epsilon_{\rm B}^{\frac{p+1}{4}}\nu^{\frac{1-p}{2}} \,
D^{-2}\,\max{\left(\theta_{\rm j},\theta_{\rm v}\right)}^{-2p}\ ,
\end{equation}
as long as the spectral regime remains $\nu_\mathrm{m}< \nu < \nu_\mathrm{c}$. 
This leads to the following expression\footnote{We neglect the effect of redshift, as the events at play here are within the GW horizon distance of first generation interferometers, i.e. with $z \leq 0.1$.} of the peak flux at $3\, \mathrm{GHz}$:
\begin{eqnarray}
F_{p,3\,\rm GHz}\ & = & 8.6\,E_{52}\,\theta_{{\rm j},-1}^2\,n_{-3}^{4/5}\,
\epsilon_{{\rm e},-1}^{6/5}\,\epsilon_{{\rm B},-3}^{4/5}\nonumber\\
& & \times\ D_{100}^{-2}\,\max{\left(\theta_{{\rm j},-1},\theta_{{\rm v},-1}\right)}^{-4.4}
\ \ {\rm mJy}\ ,
\label{eq:FP}
\end{eqnarray}
where $E_{52}=E_\mathrm{iso,c}/10^{52}$ erg, $\theta_{{\rm j},-1}=\theta_{\rm j}/0.1$ rad, 
$n_{-3}=n/(10^{-3}\ {\rm cm}^{-3})$, $\epsilon_{{\rm e},-1}=\epsilon_{\rm e}/10^{-1}$,
$\epsilon_{{\rm B},-3}=\epsilon_{\rm B}/10^{-3}$, 
$\theta_{{\rm v},-1}=\theta_{\rm v}/0.1\ {\rm rad}$
and $D_{100}=D/(100\ {\rm Mpc})$, and where we assume $p=2.2$. 
This scaling law is plotted in Fig.~\ref{fig:radiofit} (right, solid black line) and agrees well with the detailed calculation.
The expression of 
the corresponding peak time is also given by \citet{114}, assuming lateral expansion of the jet.
However, even if late observations of GRB 170817A may give some evidence for lateral expansion of the core jet (\new{for example the} temporal decay index \new{being} $\sim~-p$, as found by \citealt{80}), it is not clear if 
this 
expansion should be as strong for a core jet embedded in a sheath as for a ``naked'' top-hat jet. Therefore we also consider a limit case without lateral expansion. This only slightly affects the peak flux  
and Eq.~\ref{eq:FP} above remains a good approximation in both cases. On the other hand, it modifies the peak time, leading to:
\begin{equation}
\begin{array}{l}
t_{p, {\rm no~ex}} \sim t_b\times \left(\frac{\theta_{\rm v}}{\theta_{\rm j}}\right)^{8/3}\\
\hspace{1cm}=137\,\left({\frac{E_{52}}{n_{-3}}}\right)^{1/3}\,\theta_{v,.35}^{8/3}\ \ {\rm days}
\end{array}
\end{equation}
and
\begin{equation}
\begin{array}{l}
t_{p, {\rm ex}} \sim t_b\times \left(\frac{\theta_{\rm v}}{ \theta_{\rm j}}\right)^{2}\\
\hspace{0.7cm}=60\,\left(\frac{E_{52}}{n_{-3}}\right)^{1/3}\,\theta_{j,-1}^{2/3}\,\theta_{v,.35}^2\ {\rm days}
\end{array}
\label{eq:TP}
\end{equation}
where $t_b$ is the standard jet break time,
\begin{equation}
t_b=4.9\,(E_{52}/n_{-3})^{1/3}\,\theta_{j,-1}^{8/3}\ \ \ {\rm  days}\ \ .
\end{equation}
The ratio of the peak times without and with lateral expansion is simply
\begin{equation}
\frac{t_{p,{\rm without}}}{t_{p,{\rm with}}}\sim \left(\frac{\theta_{\rm v}}{\theta_{\rm j}}\right)^{2/3}\ . 
\end{equation}
In the following, except if mentioned otherwise, we use Eqs.~\ref{eq:FP}-\ref{eq:TP} to estimate the peak flux and peak time of the radio afterglow from a binary neutron star merger.

\begin{table*}[t]
\caption{Fiducial population model parameter distributions.}
\begin{center}
\begin{tabular}{lll}
\hline
\hline
Parameter & Symbol & Probability Distribution Function\\
\hline
Core jet isotropic equivalent kinetic energy & $E_\mathrm{iso,c}$ & Broken power law (Eq.~\ref{eq:phiE}), \\
 & & with $\alpha_1 = 0.53$, $\alpha_2 = 3.4$ \\
 & & and \new{$E_{\rm min}, E_b, E_{\rm max} = 10^{50}, 2\,10^{52}, 10^{53}~{\rm erg}$}\\

Jet half-opening angle & $\theta_{\rm j}$ & $0.1~ \rm rad \sim 5.7^\circ$\\

External medium density & $n$ & Log-normal distribution, with central value $n_0 = 10^{-3}\ \rm{cm}^{-3}$ \\
& & and standard deviation $\sigma_n = 0.75$ \\

Electron redistribution parameter & $\epsilon_{\rm e}$ & Fixed at 0.1 \\

Magnetic field redistribution parameter & $\epsilon_{\rm B}$ & Same log-normal as $n$, restricted to $[10^{-4}, 10^{-2}]$\\

Electron population spectral index & $p$ & Fixed at 2.2\\
\hline
\end{tabular}
\end{center}
\label{tab:fiducial}
\end{table*}

\subsection{Physical parameters of the jet and their intrinsic distribution}
\label{fiducial}
To generate a population of jet afterglows we have to fix the various parameters appearing in Eqs.~\ref{eq:FP}--\ref{eq:TP}. We first define a set of
``fiducial distributions'' for these parameters.
\new{These are summarized} in Tab.~\ref{tab:fiducial} and have been chosen as follows: 
  
\begin{itemize}
    \item \textit{Core jet isotropic equivalent kinetic energy $E_\mathrm{iso,c}$}: 
    we adopt a broken power law distribution, which we directly deduce 
    from the gamma-ray luminosity 
function of cosmological short GRBs, assuming a standard rest frame duration $\langle \tau \rangle=0.2$ s and a standard efficiency
$f_\gamma=0.2$ \citep{115} so that
$E_\mathrm{iso,c}\sim {L_\gamma\,\langle \tau\rangle}/ f_\gamma$, 
leading to a density of probability
\begin{eqnarray}
\phi\left(E_\mathrm{iso,c}\right) & = &\frac{1}{N}  \frac{\mathrm{d}N}{\mathrm{d}E_\mathrm{iso,c}} \nonumber\\
   & \propto & \left\lbrace\begin{array}{ll}
E_{\rm iso,c}^{-\alpha_1} & {\rm for}\ \ E_{\rm min} \leq E_{\rm iso,c} \leq E_{\rm b}\\
E_{\rm iso,c}^{-\alpha_2} & {\rm for}\ \ E_{\rm b} \leq E_{\rm iso,c} \leq E_{\rm max}
\end{array}\right.
\label{eq:phiE}
\end{eqnarray}
where $\phi$ is normalized to unity.
We fix $E_\mathrm{min}=10^{50}\, \mathrm{erg}$ and $E_\mathrm{max}=10^{53}\, \mathrm{erg}$.
Luminosity functions for short GRBs have been deduced from observations
by several groups \citep[e.g.][]{116, 105, 104}. For our fiducial model, we adopt that of \cite{104} with $\alpha_1 =  0.53$, 
$\alpha_2 = 3.4$ 
and a break luminosity $L_\mathrm{b}=2\,10^{52}\, \mathrm{erg.s^{-1}}$
leading to $E_\mathrm{b}=2\, 10^{52}\, \mathrm{erg}$.
\item \textit{Jet opening angle $\theta_{\rm j}$:} for simplicity we assume a single value $\theta_{\rm j}=0.1~{\rm rad}$, which 
appears to be 
representative 
of the typical value found by fitting the afterglow light curve of GW170817 \citep[e.g.][]{7,88}, by the VLBI observations of the remnant \citep{79} and also consistent with 
the results of prior short GRB afterglow fitting \citep{118} and short GRB/BNS merger rate comparisons \citep{67}.
\item \textit{External density $n$:} most BNS mergers are expected to occur in low density environments due to a long merger time, as assessed by the significant offsets of short GRB sources from their host galaxies \citep[e.g.][]{163, 160, 161, 162, 28}. In the case of GW170817, afterglow fitting leads typically to $n \sim 10^{-3}\, \mathrm{cm^{-3}}$, in agreement with the \new{estimation} of the HI content of the host galaxy NGC 4996 \citep{12}. 
The external densities observed in short GRBs cover the interval $10^{-3} - 1~{\rm cm}^{-3}$ \citep{28}, but this is probably biased towards high densities, which favor afterglow detection.
In this work we consider $n\sim 10^{-3}$~cm$^{-3}$ as typical 
and therefore take for the fiducial density distribution a log-normal of mean $10^{-3}~{\rm cm}^{-3}$ with a standard deviation of $0.75$.

\item \textit{Microphysics parameters:} \new{we fix the value of $\epsilon_e$ to 0.1. This is generally done in GRB afterglow modeling \citep{16, 41, 140, 117, 13}. Also, this value is adopted or found in the fitting of the afterglow of GW170817 \citep[e.g.][]{88, 17, 83, 159}, and is consistent with particle shock-acceleration simulations \citep{SS2011}}.
Similarly, we fix $p$ to 2.2, which is also a common value used in GRB afterglow models and is in agreement with shock-acceleration theory in the relativistic regime \citep[e.g.][]{144}. In the case of GW170817 this value can be directly measured  from multi-wavelength observations \citep{80, 5, 88}. 
Finally, more diversity is generally expected for $\epsilon_{\rm B}$.
We assume a log-normal distribution similar to the one adopted for $n$ with the additional constraint that $\epsilon_{\rm B}$ is restricted to the interval $[10^{-4},10^{-2}]$.
\end{itemize}
 

In Sec.~\ref{sec:variations}
we consider possible alternatives to these fiducial distributions. In particular, we discuss the impact of the large uncertainties on the distribution of kinetic energy, $\phi\left(E_\mathrm{iso,c}\right)$, either due to the difficulty to measure the luminosity function of short GRBs, or to our simplifying assumptions regarding the duration and the efficiency of the prompt GRB emission. We also explore different values of the jet opening angle $\theta_{\rm j}$. 
As for the external density, we study the effect of a change in the mean circum-merger density of the population of mergers.

\section{From the intrinsic to the observed population: GW+radio joint detection}
\label{criteria}

\subsection{GW detection criterion}
\label{sec:detGW}
The signal-to-noise ratio (SNR, denoted by $\rho$) of a GW inspiral signal in a single LIGO-Virgo-type interferometer can be written as $\rho^2 = \frac{\Theta^2}{D^2}  \mathcal{M}^{5/3} S_I $ \citep{65}, where $D$ is the luminosity distance to the binary, $\mathcal{M}$ is the chirp mass of the binary, $S_I$ is a quantity depending only the sensitivity profile of the interferometer, and $\Theta^2$ represents the dependence of the signal to the binary sky position $(\theta, \phi)$ and orientation $(\theta_{\rm v}, \xi)$ with respect to the plane of the interferometer. Again, $\theta_{\rm v}$ is the angle of the line-of-sight to the binary polar direction, which we will also suppose is the jet-axis direction. \new{The function} 
$\Theta^2$ admits a global maximum of $\Theta^2_M = 16$ corresponding to an optimally positioned and oriented source (binary at the zenith and polar axis orthogonal to the instrument's arms). This optimal binary can be detected out to a distance known as the \textit{horizon} $H$ of the instrument, which depends on the SNR threshold for detection, taken to be $\rho_0 = 8$ for the LVC network. We may thus rewrite the SNR in the following manner:
\begin{equation}
\rho^2 = \rho_0^2 \frac{\Theta^2}{\Theta_M^2}\frac{H^2}{D^2}\, .
\end{equation}
In a full population study, one would have to draw all four angles and evaluate this criterion in every case. We choose to reduce the number of parameters to two: $D$ and $\theta_{\rm v}$, as these are the ones relevant to the afterglow. We must thus review this criterion by averaging $\Theta^2$ on the sky-position $(\theta, \phi)$ and polarization angle $\xi$. This is readily done from the expression given in \cite{65} (Eq.~3.31) and it is found analytically that:
\begin{equation}
\braket{\Theta^2}_{\xi, \theta, \phi} = \frac{1}{4\pi 2\pi}\int \mathrm{d}\xi\, \mathrm{d}\Omega\,  \Theta^2 = \frac{4}{5}\left( 1 + 6\cos^2 \theta_{\rm v} + \cos^4 \theta_{\rm v} \right)\, .
\end{equation}

Hence, we \new{shall} use the following criterion to determine those binaries of inclination $\theta_\mathrm{v}$ and distance $D$ which are detectable in GW on average in the sky: 
\begin{equation}
\label{crit}
\braket{\rho^2}_{\xi, \theta, \phi} > \rho_0^2\, ,
\end{equation}
which is:
\begin{equation}
\label{crit2}
\sqrt{\frac{1 + 6\cos^2 \theta_{\rm v} + \cos^4 \theta_{\rm v}}{8}} > \frac{D}{\bar{H}}\, ,
\end{equation}
where we have denoted $\bar{H} = \sqrt{\frac{2}{5}} H \sim H / 1.58$ the \textit{sky-position-averaged} horizon.

Averaging $\Theta^2$ on inclination angle $\theta_{\rm v}$ and imposing a SNR threshold as in Eq.~\ref{crit} would lead to the simple detection criterion of $D < R$, where $R$ is the \textit{range} of the instrument, i.e. the maximum distance to which a binary can be detected on average in \textit{sky-position} and in \textit{orientation}. The range is linked to the horizon by $H = 2.26 R$.

The criterion of Eq.~\ref{crit2} is valid for a detection using a single instrument. Instead, GW detection by the interferometer network is based on multi-instrument analysis, and is thus more complicated than that described by our criterion. Furthermore, as it was illustrated in the case of GW170817, true joint GW+AG detection requires the pin-pointing of the source \new{through the detection of the kilonova counterpart\footnote{This \new{was not achieved} in the recent merger candidates S190425z and S190426c \citep{GCN1, GCN2}, see Sec.~\ref{sec:discussion} for more discussion on this.}}, which in turn involves a small enough localization map from the GW data, and our criterion should incorporate this. We choose a simple localization criterion by supposing that a source is localized if it is detectable (once again on average) by the two most sensitive interferometers of the network. Thus our detection-localization criterion is that of Eq.~\ref{crit2}, with the horizon of the second most sensitive instrument of the network, i.e. the LIGO-Hanford instrument. Tab.~\ref{horizon} reports the corresponding values taken for our study for the LVC O2, O3 runs and for the design interferometers.

In this framework, \new{it can be shown that assuming homogeneity of the sources within the horizon and isotropy of the binary polar direction,} the mean viewing angle of GW-detected events is $\sim 38^\circ$ and the fraction of GW-detected events among all mergers is $\sim 29\%$, both regardless of the horizon value. This \new{latter fraction} is thus an absolute maximum to the fraction of mergers to be jointly detected in both radio and GW channels.

\begin{table}[!h]
\caption{Horizons used for the GW detection-localization criterion in our study. They are deduced from the published BNS \textit{ranges} of the instruments \citep{119} by $H=2.26 R$ and considering the value for LIGO-Hanford (see text for details of this choice). }
\begin{center}
\begin{tabular}{lll}
\hline
\hline
LVC Run & $H$ (Mpc) & $\bar{H} = H \times \sqrt{{2}/{5}}$ (Mpc) \\
\hline
O2 & 136 & 85.8 \\
O3 & 226 & 143 \\
Design & 429 & 272\\
\hline
\end{tabular}
\end{center}
\label{horizon}
\end{table}

\begin{figure*}[!t]
\resizebox{\hsize}{!}{\includegraphics{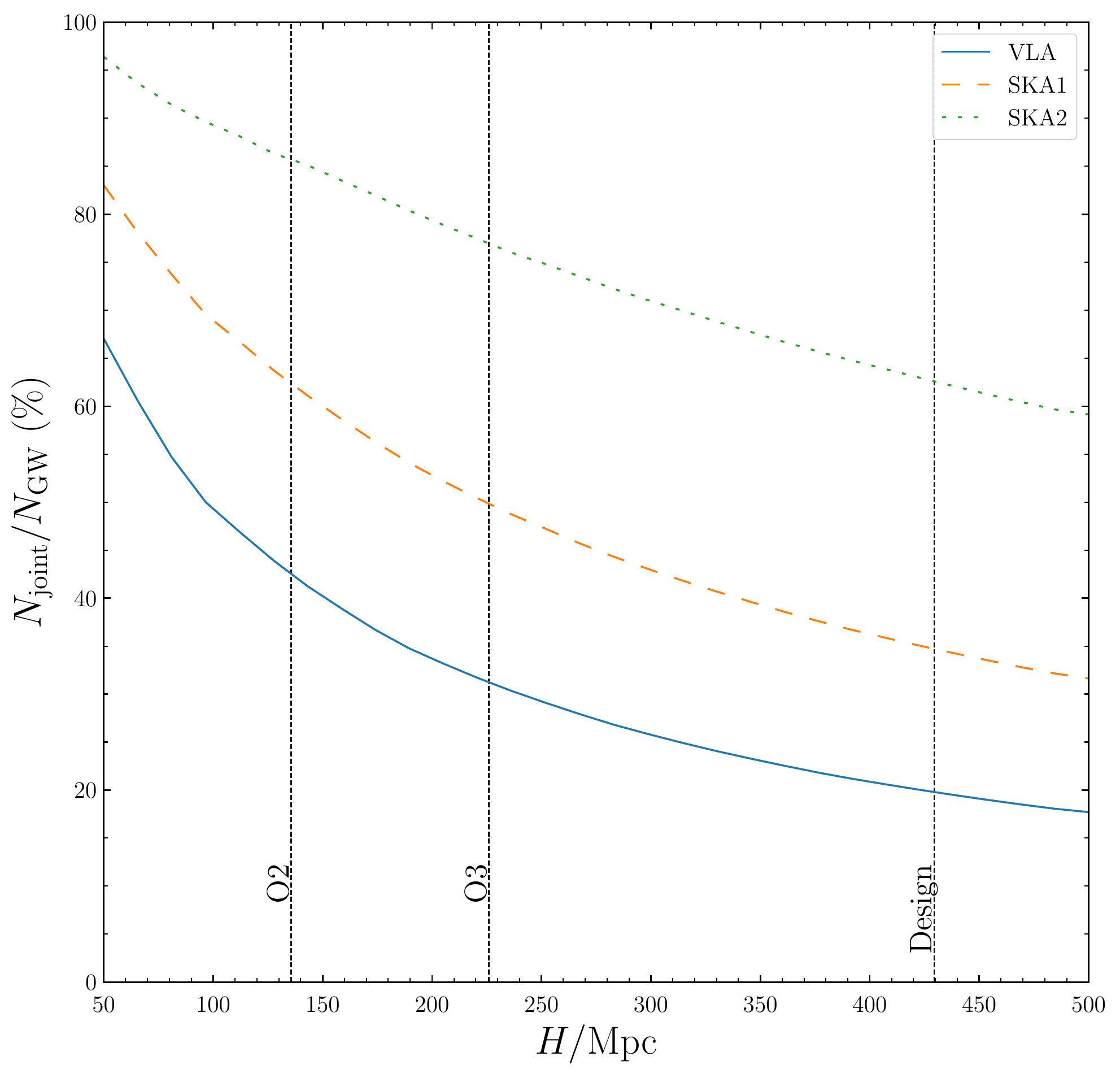}\includegraphics{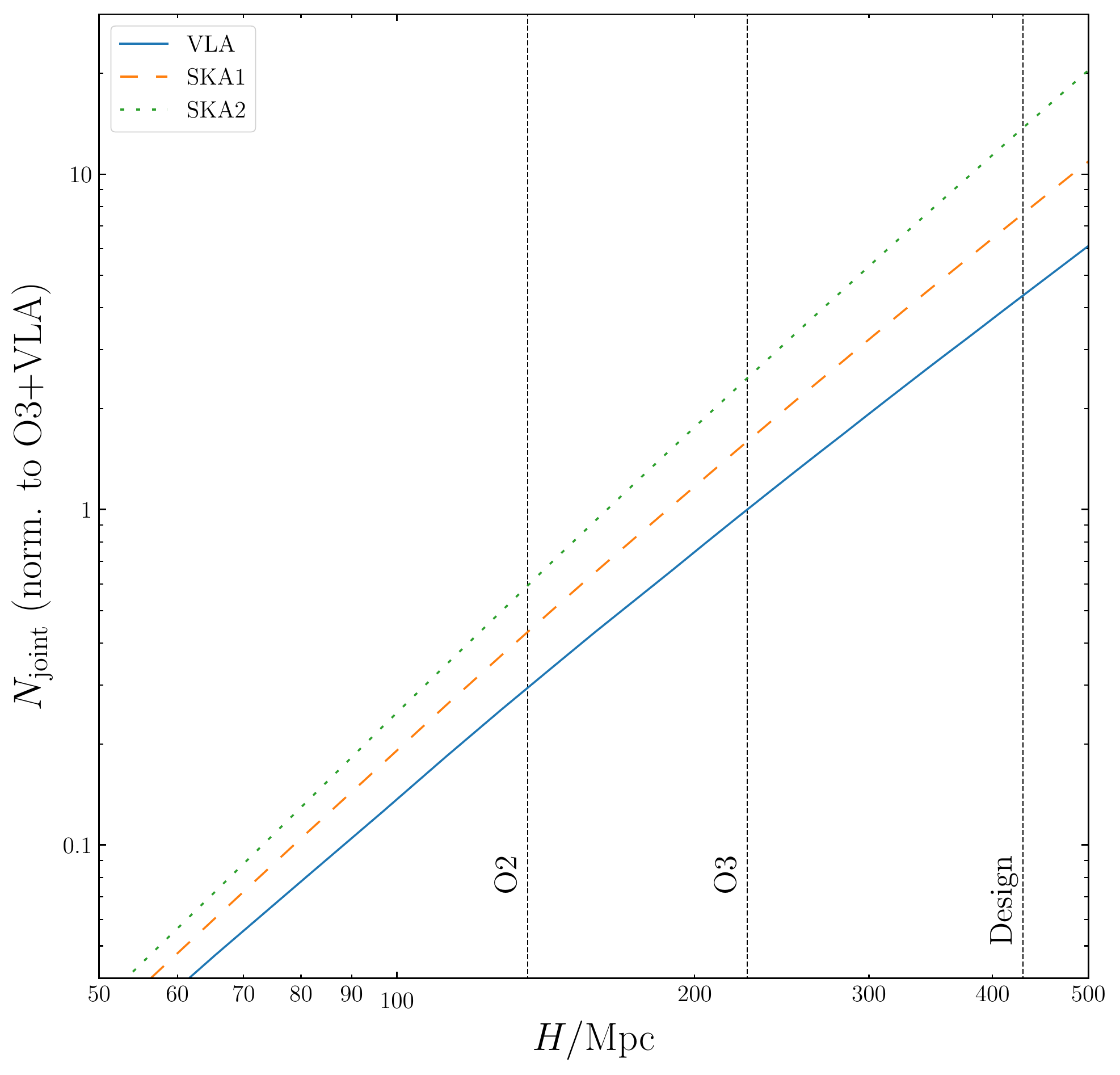}}
\caption{\textit{Left}: \new{Detectable} fraction of radio afterglows among gravitational wave events as a function of the \textit{horizon} distance $H=1.58\bar{H}$. 
\textit{Right}: Expected number of \new{joint} detections normalized to the case of the \new{O3+VLA configuration}. In both panels the full (resp. dashed, resp. dotted) lines correspond \new{to the VLA (resp. SKA1, resp. SKA2/ngVLA) being the limiting radio facility}.}
\label{fig:fracH}
\end{figure*}

\subsection{Radio detection criterion}
The detection criterion in the radio band is simply that the 3~GHz peak flux as determined by our models be larger than the sensitivity of the radio array available for follow-up at the time considered, which we will denote by $s$. We note that this is a criterion for detectability rather than detection. True detection only occurs if observations are pursued for long enough after the GW merger signal, as the flux rises to its maximum. We also note that events which are marginally detectable will provide only poor astrophysical output because the inference of jet parameters requires the fitting of an extended portion of the radio afterglow light-curve.

We will take \new{three} typical radio sensitivities reflecting the present and future capabilities of \new{radio} arrays: (i) the Karl Jansky Very Large Array (VLA) at
\new{$s=15\, \mathrm{\mu Jy}$}, 
(ii) \new{the phase I of the Square Kilometer Array (SKA1, \citealt{SKA}), at $s=3\, \mathrm{\mu Jy}$,}
and (iii) \new{phase II of the SKA (SKA2) and the Next Generation VLA (ngVLA, \citealt{ngVLA}), both at $s=0.3\, \mathrm{\mu Jy}$}.
The combination of this radio detection criterion with that in the GW domain (Eq.~\ref{crit2}) constitutes \new{our} criterion for joint detection of mergers.

\new{In the sequel, we will refer to a GW and radio jointly detectable population by using the names of the corresponding LVC run and limiting radio facility, as in ``the joint population of the O3+VLA configuration''. The corresponding GW interferometer horizons and radio sensitivities can be found respectively in Tab.~\ref{horizon} and in the paragraph above.}

\section{Results: detection rates and properties of the detected population}
\label{sec:results}

\subsection{Fiducial model}

In this section, we describe the population of events detected jointly in the GW and radio domains.
For this purpose, we use a Monte Carlo approach where we simulate $N>10^6$  binary systems within the sky-averaged horizon $\bar{H}$ using the parameter distributions of
the fiducial model given in Sec.~\ref{fiducial}.
Applying the detection criteria described in Sec.~\ref{criteria}, we obtain $N_\mathrm{GW}$ systems detected by the GW interferometer network and $N_\mathrm{joint}$ systems which are also detectable by radio telescopes through their afterglows. Then we study the properties of the sample of joint GW+radio detectable events.

\subsubsection{Rate of joint GW+Radio detectable events}

\label{sec:rates}

Starting from the fraction $f_{\rm GW}~=~N_\mathrm{GW}/N~\sim 29\%$ of binary systems detected by the LVC (Sec.~\ref{sec:detGW}), we will now
estimate which fraction $N_\mathrm{joint}/N_\mathrm{GW}$ of those events also produce a detectable radio afterglow.

\new{This fraction is} shown in Fig.~\ref{fig:fracH} (left) \new{as a function of the GW horizon and for the VLA, SKA1 and SKA2/ngVLA limiting sensitivities.}
In the design configuration of the LVC network,
\new{75\% (resp. 2 time) more events are detectable by SKA1 (resp. SKA2/ngVLA) than by the VLA.}

The number of \new{joint event} detections \new{normalized to the O3+VLA configuration} is represented in Fig.~\ref{fig:fracH} (right). 
The number of joint detections behaves approximately as $H^\alpha$, with $\alpha < 3$ because of the reduction of the radio detection efficiency when the distance increases (Fig.~\ref{fig:fracH}, left). We find $\alpha\simeq 2.4$ (resp. $2.6$, resp. $2.8$) for the \new{VLA} (resp. \new{SKA1}, resp. \new{SKA2/ngVLA}).


As illustrated in Fig.~\ref{fig:fracH} (left) we find that the fraction of detected events decreases from \new{43\%} 
(O2) to %
\new{31\%}
(O3) with \new{the VLA} limiting sensitivity. 
However the absolute number of detections will increase (Fig.~\ref{fig:fracH}, right).
With the design+\new{SKA1,} 
a fraction of \new{$35\%$} 
of GW events can lead to a radio detection.
\new{With the much higher sensitivities that may be reached after 2030, more than a half of GW events may become detectable in radio. Indeed we find a fraction of 63\% of detectable joint events in the design+SKA2/ngVLA configuration. }


For completeness, the number of GW and jointly detected events per continuous year of GW network operation for future detector configurations can be found in Tab.~\ref{numbers}, where we have used the horizons of Tab.~\ref{horizon} and our fiducial model. 

\subsubsection{Distance and viewing angle}
\label{sec:4.1.2}
\begin{table*}[!t]
\caption{Number of \textit{detectable} events per continuous year of GW network operation. The expected rate of total GW events is taken from \cite{119} and values for joint events are derived using our fiducial population model. The uncertainties here are due to the uncertainty on the merger rate in the local Universe derived from GW observations of the LVC O1 and O2 runs. The additional uncertainties related to the population model will be discussed in Sec.~\ref{sec:variations}.}
\begin{center}
\begin{tabular}{llllll}
\hline
\hline
LVC Run & \multicolumn{2}{c}{Radio Configuration}  & GW Events & Joint Events & \new{Fraction of detectable events}\\
 &Instrument &$s$ ($\mu$Jy) &  $N_\mathrm{GW}$ & $N_\mathrm{joint}$ & \new{(assuming fiducial model)}\\
\hline
O3 & \new{VLA} &\new{$15$} &  $9^{+19}_{-7}$ & $3^{+6}_{-2}$  & \new{31.4\%}\\
Design & \new{VLA} &\new{$15$} &  $21^{+44}_{-16}$ & $4^{+10}_{-4}$ &\new{19.8\%}\\
Design & \new{SKA1} &\new{$3$} &  $21^{+44}_{-16}$ & $7^{+18}_{-7}$ &\new{34.7\%}\\
Design & \new{SKA2/ngVLA} &\new{$0.3$} &  $21^{+44}_{-16}$ & $13^{+33}_{-13}$ & \new{62.5\%}\\
\hline
\end{tabular}
\end{center}
\label{numbers}
\end{table*} 

\begin{figure*}
\resizebox{\hsize}{!}{\includegraphics{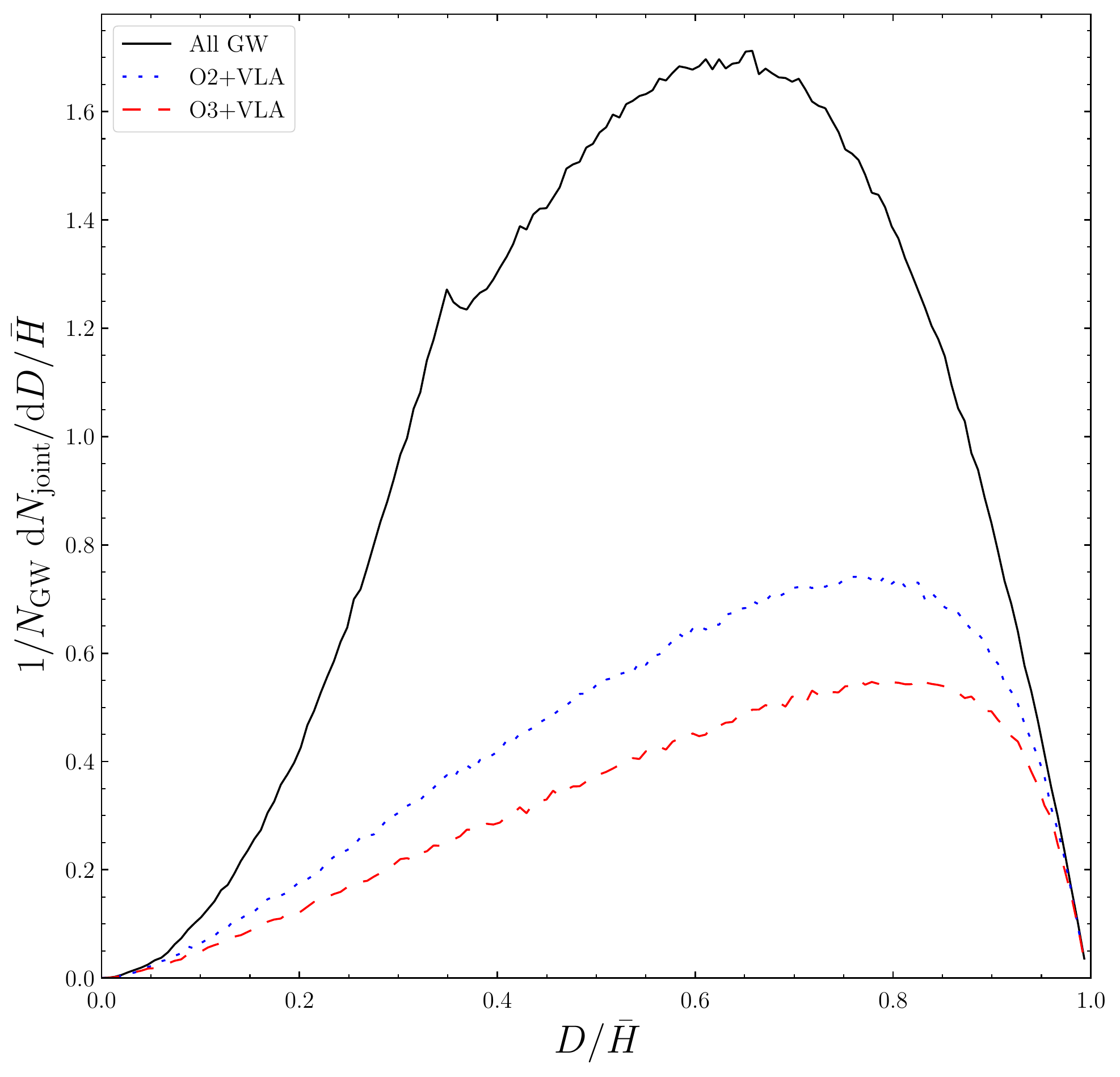}\includegraphics{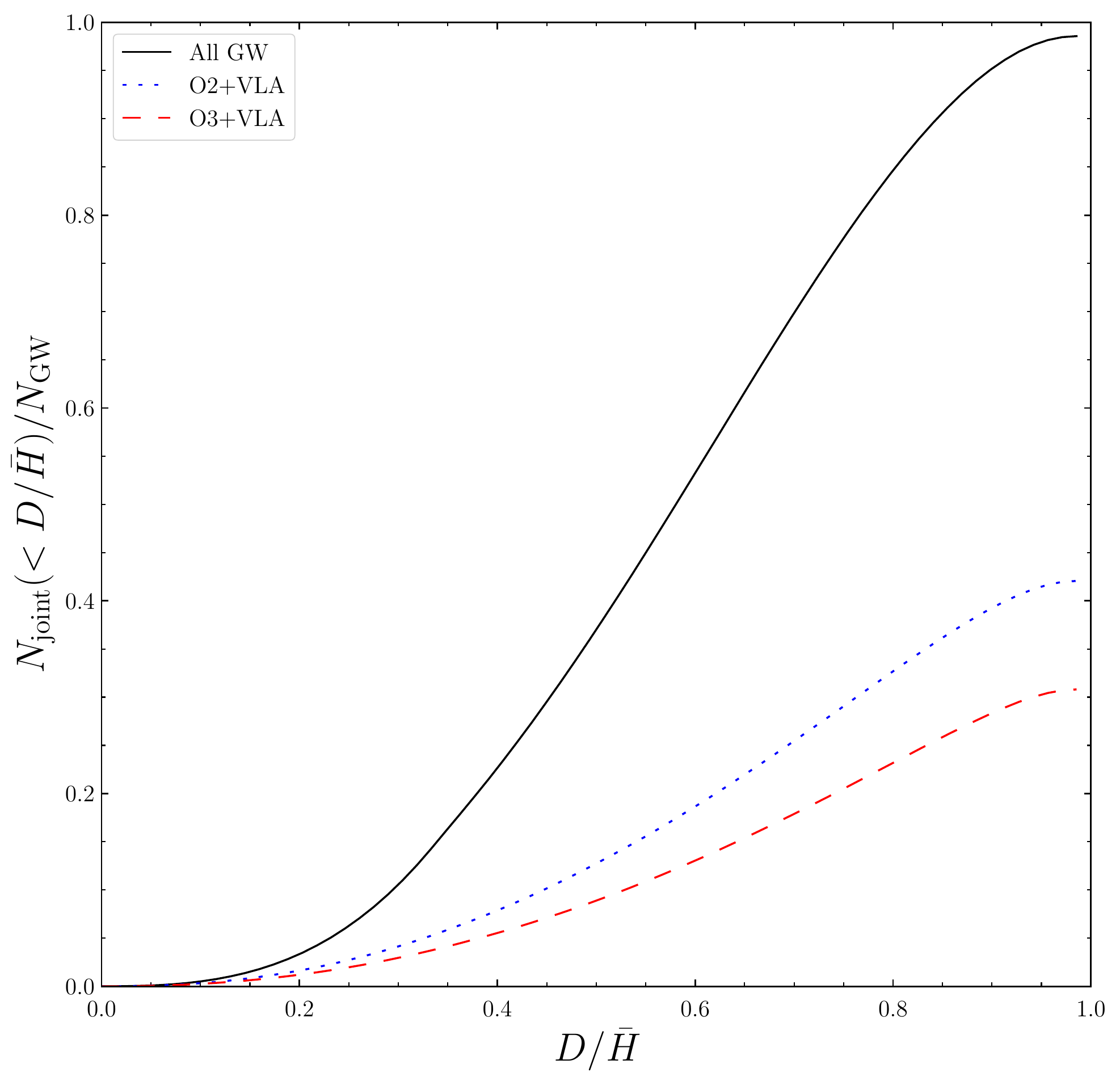}}
\caption{\textit{Left}: Differential distribution of the distances (normalized to the sky-position-averaged horizon) of events in the \new{total GW population} (full line), and jointly in the O2+VLA (dotted line) and O3+VLA (dashed line) configurations. \textit{Right}: Cumulative distribution of the same events. \new{Distributions are normalized to the fraction of jointly detected events among all GW-detected events, see text for details.}}
\label{fig:distance}
\end{figure*}

The distributions \new{of} distances are shown in Fig.~\ref{fig:distance}
for the O2+VLA and O3+VLA configurations, as well as of the entire GW population. \new{In this figure, the distributions are not normalized to unity so as to show the decreasing of the fraction of joint event as the horizon increases.}
It can be seen that as a result of the GW and radio detection thresholds, sources are progressively lost when the distance increases so that this distribution exhibits a near linear increase (as opposed to the $\mathrm{d}N/\mathrm{d}D\propto D^2$ of the intrinsic homogeneous population).
Up to $D=\bar{H}/\sqrt{8}\simeq 0.35 \bar{H}$, all events are detected in GW, regardless of their orientation\footnote{After this distance, events are detected in GW only if $\theta_{\rm v} \leq \theta_{\rm max}$ with $\cos\theta_{\rm max} = \sqrt{-3 + \sqrt{8 + 8 D^2/\bar{H}^2}}$. The differential distribution of distances \new{for the GW events} thus transitions from $\propto D^2$ before $\bar{H}/\sqrt{8}$ to a non-quadratic form afterwards, producing the small peak seen in Fig.~\ref{fig:distance}. This of course is only the consequence of 
our simplified GW sky-averaged detection criterion (Eq.~\ref{crit2}).} and the only selection is due to the radio sensitivity. 
Near the horizon, the maximum viewing angle allowing GW-detection is $\theta_\mathrm{max} \propto \sqrt{1 - D/\bar{H}}$, which produces a strict decrease of event density.

\begin{figure*}
\resizebox{\hsize}{!}{\includegraphics{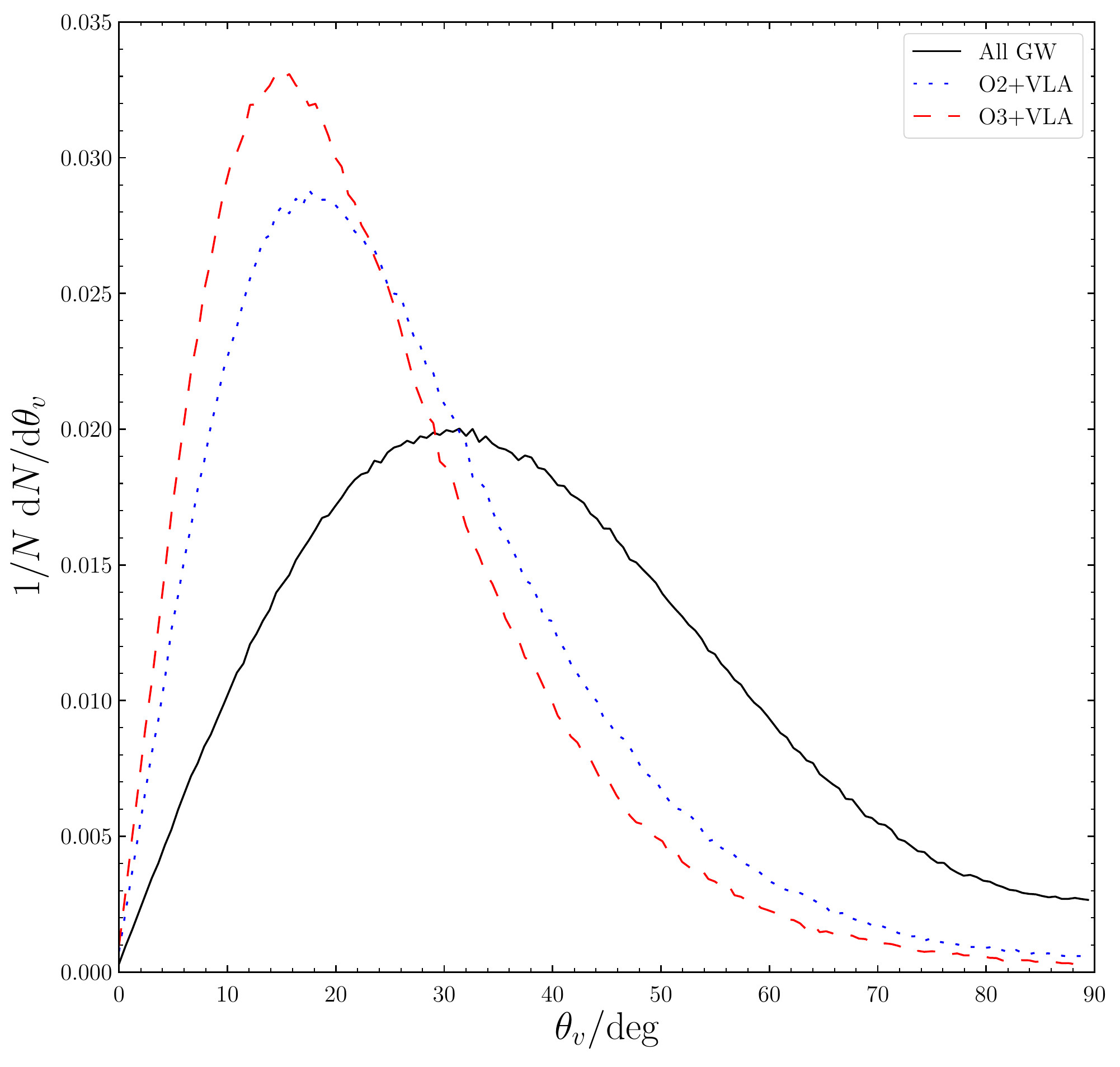}\includegraphics{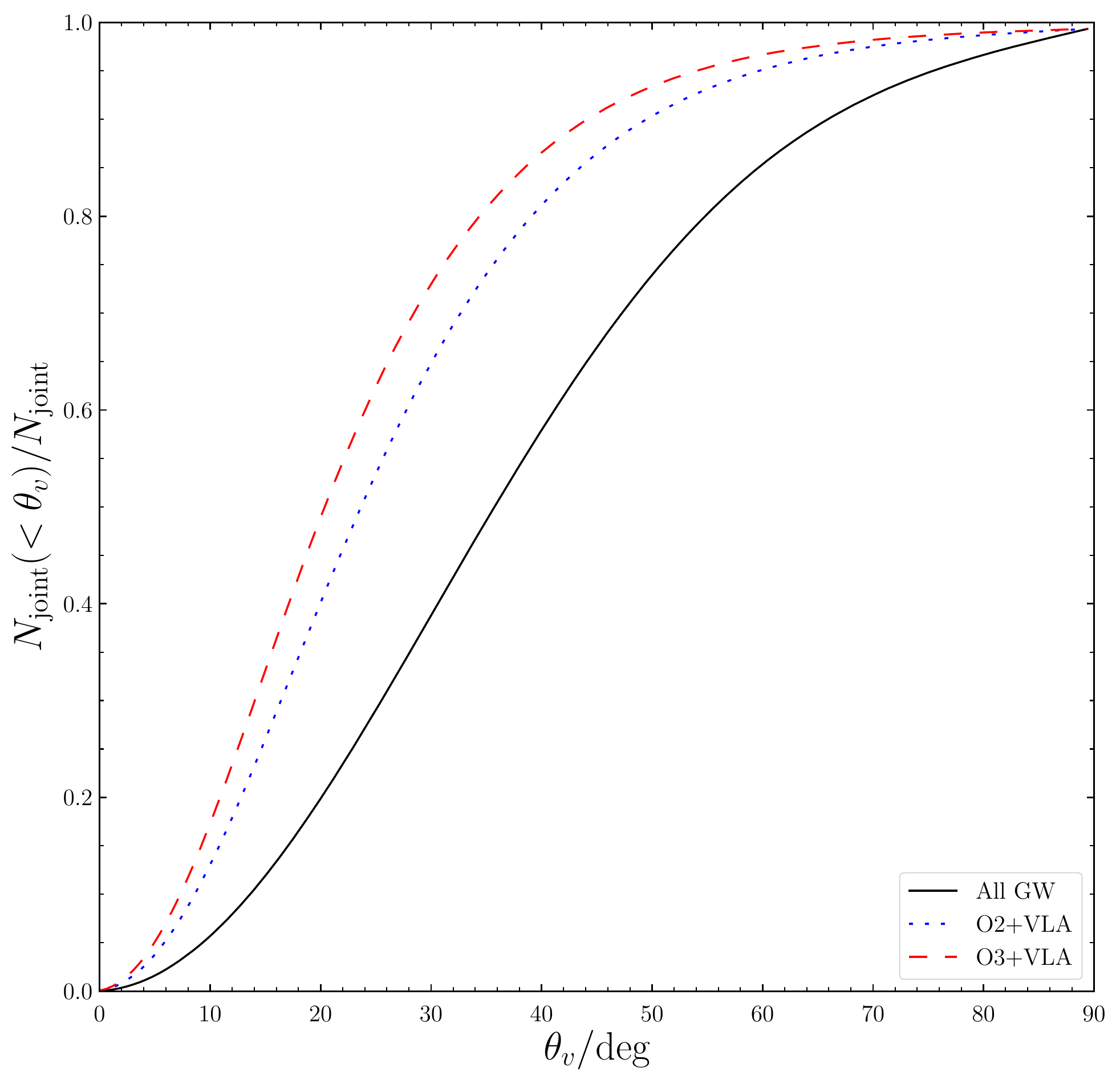}}
\caption{Same as Fig.~\ref{fig:distance}, for the viewing angle. Distributions are normalized.}
\label{fig:angle}
\end{figure*}

\begin{figure*}
\resizebox{\hsize}{!}{\includegraphics{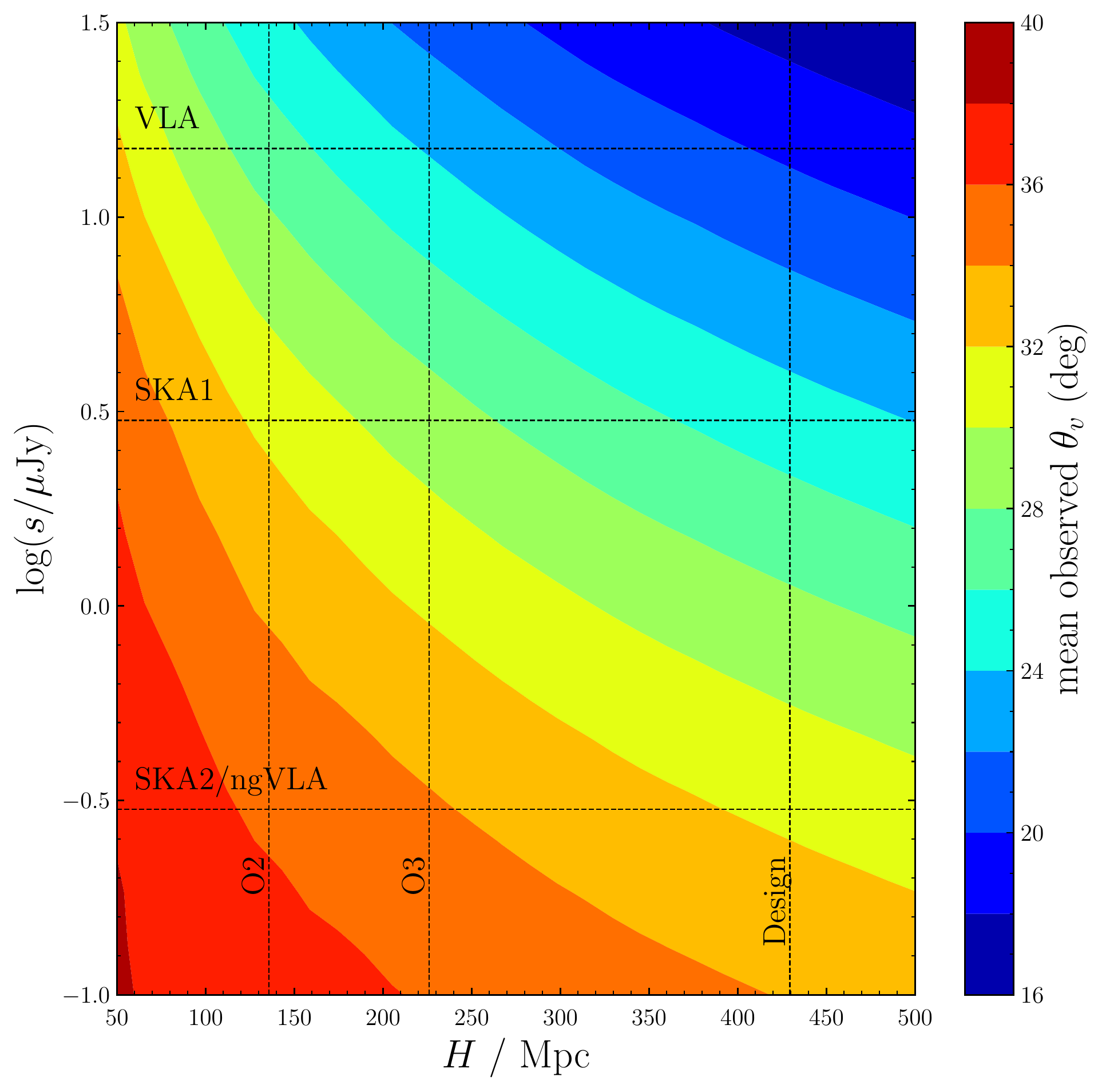}\includegraphics{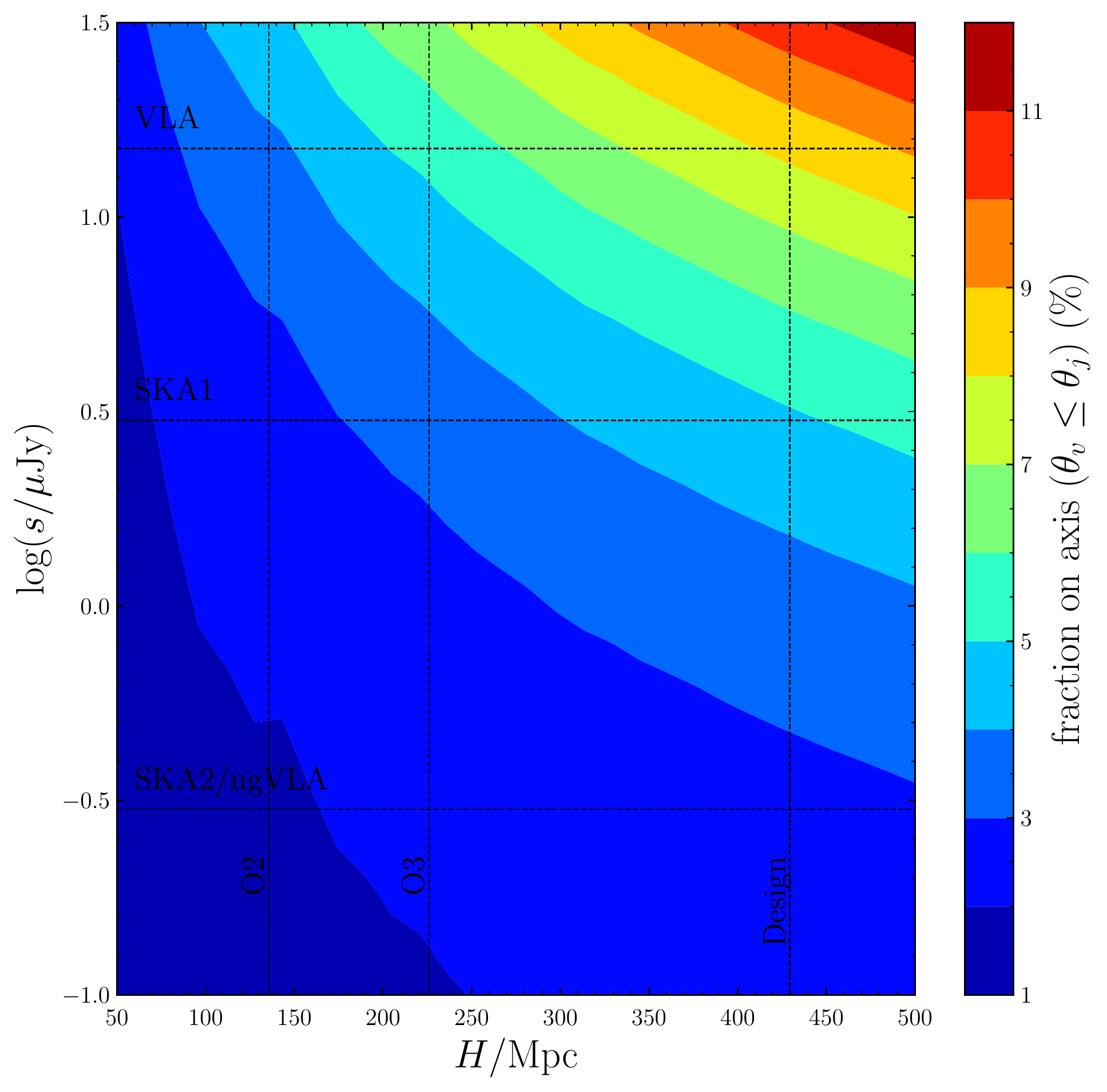}}
\caption{\textit{Left}: Mean viewing angle of the jointly detected events, as a function of the \textit{horizon} and the radio sensitivity $s$. The values of $H$ for O2, O3 and design instruments as well as $s$ for VLA, SKA1 and SKA2/ngVLA are indicated in dashed lines. 
\textit{Right}: Fraction of on-axis events (defined by $\theta_\mathrm{v}\le \theta_\mathrm{j}$) among jointly detected events in the same $H$--$s$ diagram.}
\label{fig:thetav}
\end{figure*}

Fig.~\ref{fig:angle} shows the distribution of the viewing angles. \new{For the joint events,}
the peak of the distribution takes place at small viewing angles $\theta_{\rm v}\sim 15-20^{\circ}$
\new{as a result of the rapid decline of the peak flux with angle ($F_{p}\propto \theta_{\rm v}^{-2p}$),} 
and about 39\% (resp. 47\%) of the events are seen with a viewing angle 
$\theta_{\rm v}<20^\circ$ for O2+VLA (resp. O3+VLA), the angle with which the GW 170817 event was \new{likely} seen. 
As shown in Fig.~\ref{fig:thetav} (left), the mean viewing angle of the jointly observed events strongly depends on the GW horizon and the radio sensitivity. It appears that even if the radio sensitivity is not largely improved while the GW interferometers reach their design horizons, there will remain a significant number of events seen off-axis. The increasing and decreasing phases of these events' afterglows will allow to better study the jet structure which is more revealed by off-axis events. 
Fig.~\ref{fig:thetav}~(right) shows the fraction of on-axis events ($\theta_\mathrm{v}\le \theta_\mathrm{j}$) within the joint detections. It increases from 4.0\% (O2+VLA) to 5.3\% (O3+VLA). 

\subsubsection{Peak times, peak fluxes and synchrotron spectral regime}
\begin{figure*}
\resizebox{\hsize}{!}{\includegraphics{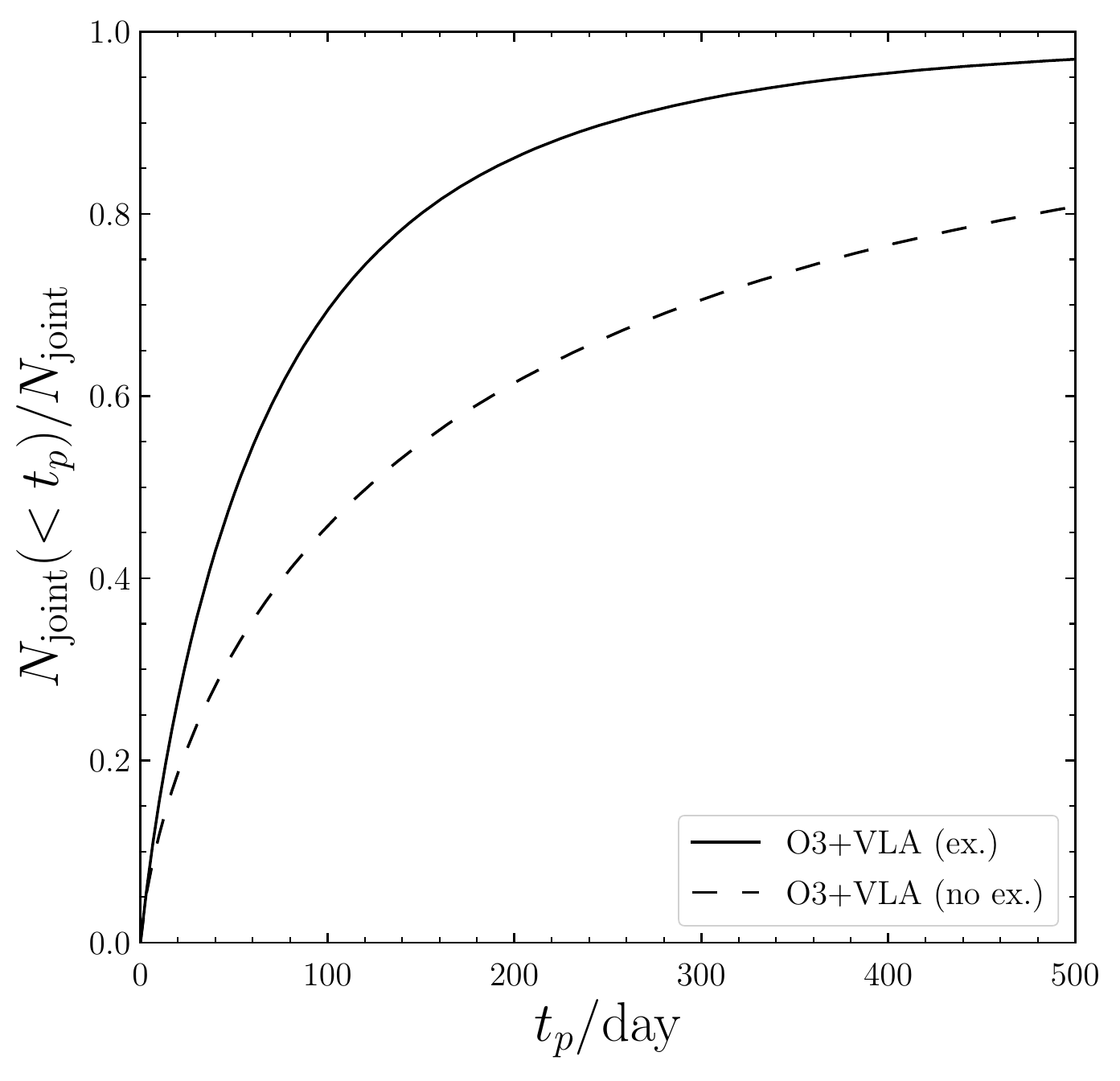}\includegraphics{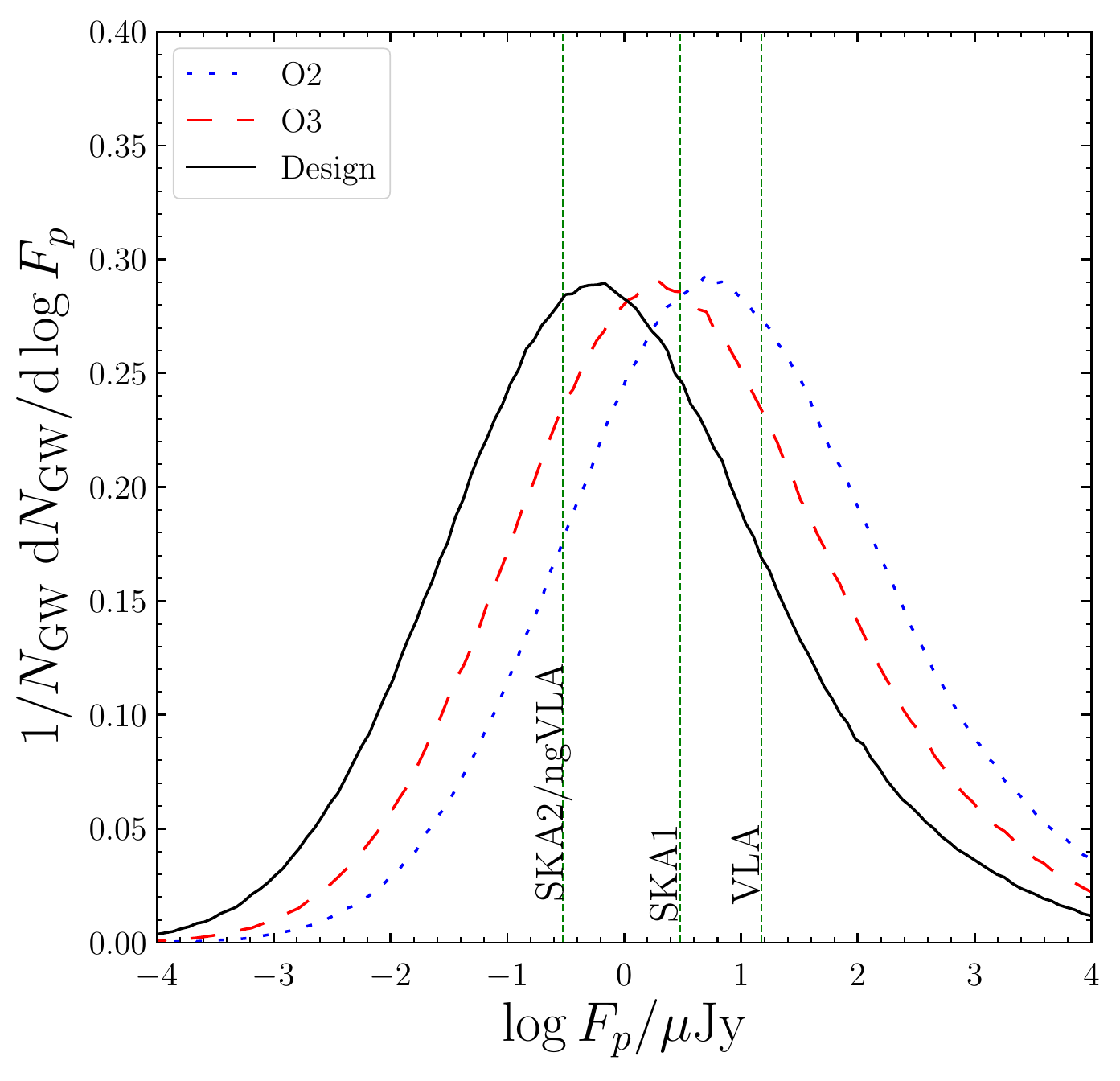}\includegraphics{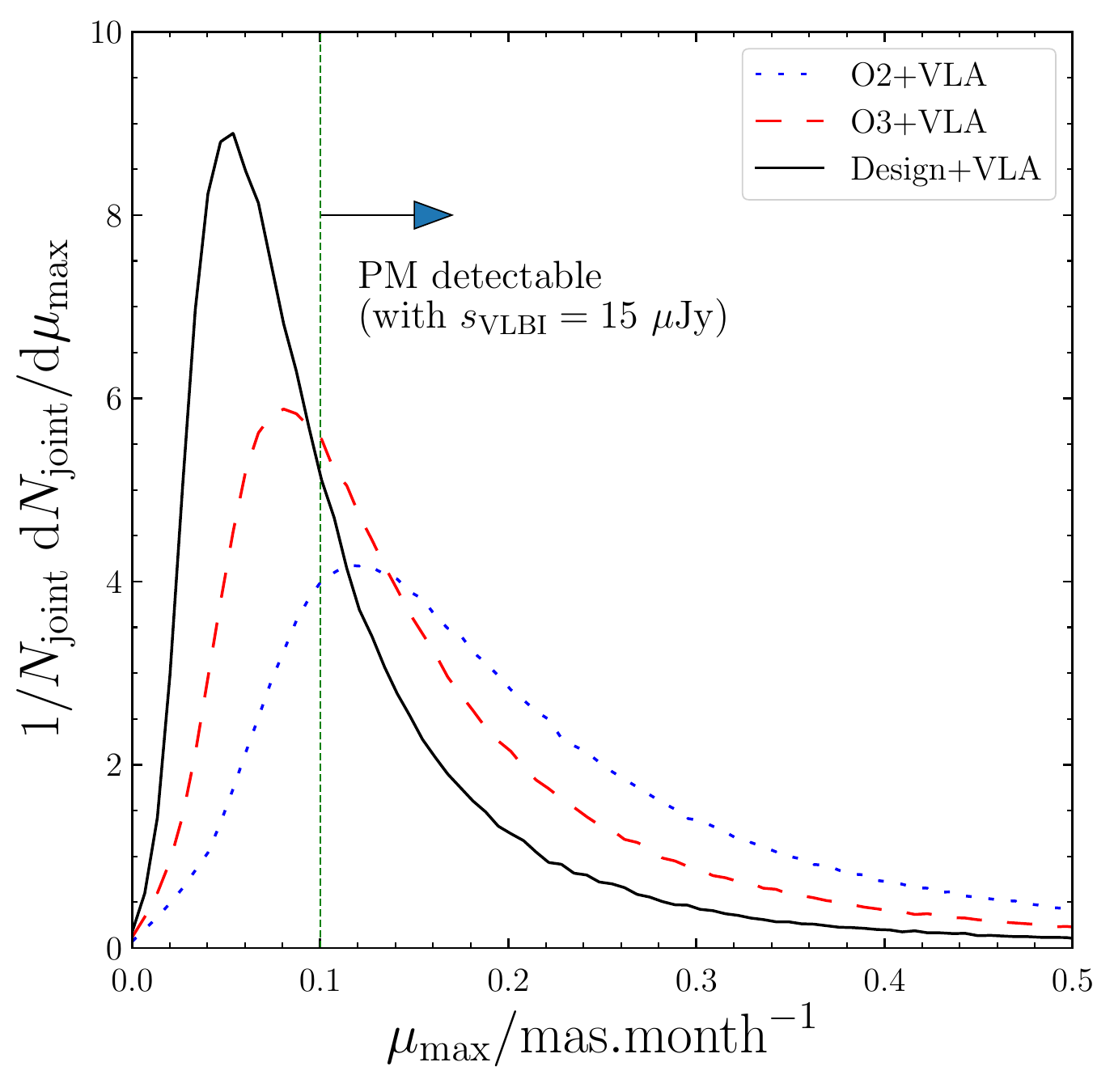}}
\caption{\textit{Left}: Cumulative distributions of the peak times of the radio afterglow for jointly detected events in the O3+VLA configuration, assuming a jet with (solid line) or without (dashed line) lateral expansion. \textit{Middle}: Differential distribution of the peak fluxes of GW-detected events assuming the horizon for the O2 (dotted line), O3 (dashed line) and design (full line) configurations of the GW interferometers. The radio sensitivities for VLA, SKA1 and SKA2/ngVLA are indicated for comparison. \textit{Right}: Differential distribution of the maximum proper motion for jointly detected events assuming the O2+VLA (dotted line), O3+VLA (dashed line), design+SKA1 (full line) configurations. The dotted vertical line shows the lower limit for detection of the proper motion.}
\label{fig:TPFP}
\end{figure*}

The distributions of peak times and peak fluxes \new{in the O3+VLA configuration} are represented in Fig.~\ref{fig:TPFP}. 
The fraction of events with a peak time smaller than 150 days (as observed in GRB170817A) \new{is} 55\% without jet expansion \new{and} 81\% with lateral expansion. 
The distribution in peak flux is shown for all sources which are detected in gravitational waves within the sky-position-averaged horizon of O2, O3 and design instruments. It appears clearly that for the present VLA sensitivity, most radio afterglows cannot be detected. This explains why improving the sensitivity of the future ngVLA has such an impact on the joint detection rates, as shown in Fig.~\ref{fig:fracH}. 

The scaling law used to compute the radio emission at the peak assumes that the observing frequency remains in the same spectral regime of the slow cooling synchrotron spectrum, i.e. $\nu _\mathrm{m}< \nu < \nu_\mathrm{c}$, and above the absorption frequency. Using our more detailed calculation of the radio afterglow from the core jet (Eq.~\ref{eq:FluxApp} and text thereabove), we could check that this condition was fulfilled for the bulk of the population. However, for mergers in high density environments (larger than $10~{\rm cm}^{-3}$), this condition is no longer met as the absorption frequency $\nu_a$ and the injection frequency $\nu_i$ may be larger than the radio frequency at early times. For an event with $n = 10~{\rm cm}^{-3}$ and $\theta_{\rm v}\sim 40^\circ$ (the mean viewing angle of the GW-detected population), the radio frequency typically meets the injection break at around 30 days (resp. \new{before 10} days). In this case, chromatic lightcurves are expected.

\subsubsection{Proper motion}
\label{sec:4.1.4}
Another independent 
observable is the proper motion of the remnant, which can be readily accessible to VLBI measurements as illustrated in the case of GRB170817A \citep{79, 110}. \new{However, a VLBI network can only measure the proper motion of a source if the total displacement of the source over the time it is detectable by the array is larger than the angular resolution of the network. For our purposes, we shall simplify this criterion and adopt a dual condition on the flux ($F_p$ > $s_{\rm VLBI}$) \textit{and} on the instantaneous proper motion $\mu_{\rm max}$ (requiring that $\mu_{max}$ > $\mu_{lim}$) of the remnant at the peak of the afterglow. Here, we will adopt a limiting proper motion of $\mu_{\rm lim}=3$ $\mu$as/day, \new{in coherence with} the uncertainties quoted in \cite{79, 110}}

An upper limit of the proper motion is obtained
assuming that the core jet contribution dominates at the peak of the
radio light curve, and that the shocked region at the front of this core jet constitutes the visible remnant of the merger.
In first approximation, afterglows from jets peak when the observer enters the focalization cone of the radiation, i.e. $\Gamma \sim 1/\theta_{\rm v}$. For an off-axis observer, we therefore estimate the maximum proper motion to be:
\begin{equation}
\label{pm}
\mu_{\rm max} = \frac{c \beta_\mathrm{app}}{D}
=\frac{c}{D}\,f(\theta_{\rm v})
\end{equation}
with
\begin{equation}
f(\theta_{\rm v}) = \frac{\sqrt{1 - \theta_{\rm v} ^2 }\sin \theta_{\rm v}}{1 - \sqrt{1 - \theta_{\rm v} ^2} \cos \theta_{\rm v}}\, .
\end{equation}

Thus, for a given value of the viewing angle, there is a maximum distance beyond which the measurement of the proper motion is not possible. It is given by
\begin{equation}
D_{\rm max}(\theta_{\rm v})=\frac{c}{\mu_{\rm lim}}\,f(\theta_{\rm v})\sim50\,f(\theta_{\rm v})\ \ {\rm Mpc}\, ,
\label{eq:dmax}
\end{equation}
. 

The distribution of $\mu_{\rm max}$ is shown in Fig.~\ref{fig:TPFP} \new{(right) for O2, O3 and design-level detectable GW events respecting the VLBI flux criterion}. Here, we take $s_{\rm VLBI} = 15~\mu{\rm Jy}$, which represents a limiting SNR of $\sim~3$ with the level of noise reported in \cite{79,110} for the e-MERLIN and High Sensitivity Array networks.
\new{Within the} O2+VLA (resp. O3+VLA, resp. design+VLA) populations the measurement of the proper motion is possible in only 79\% (resp. 64\%, resp. 42\%) of the \new{jointly-detectable events}. 

\begin{figure*}[!ht]
\begin{center}
\resizebox{0.9\hsize}{!}{\includegraphics{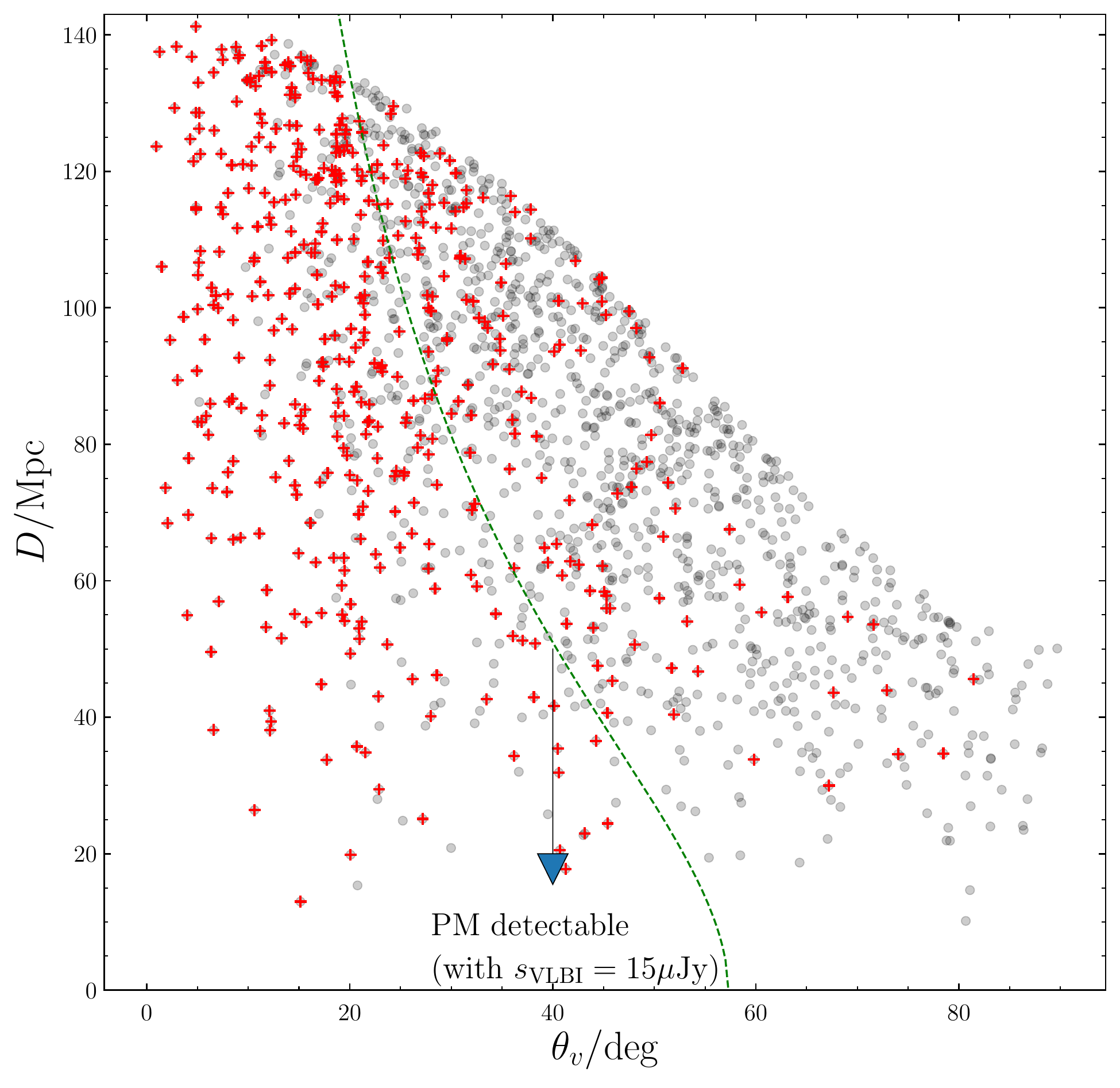}
\includegraphics{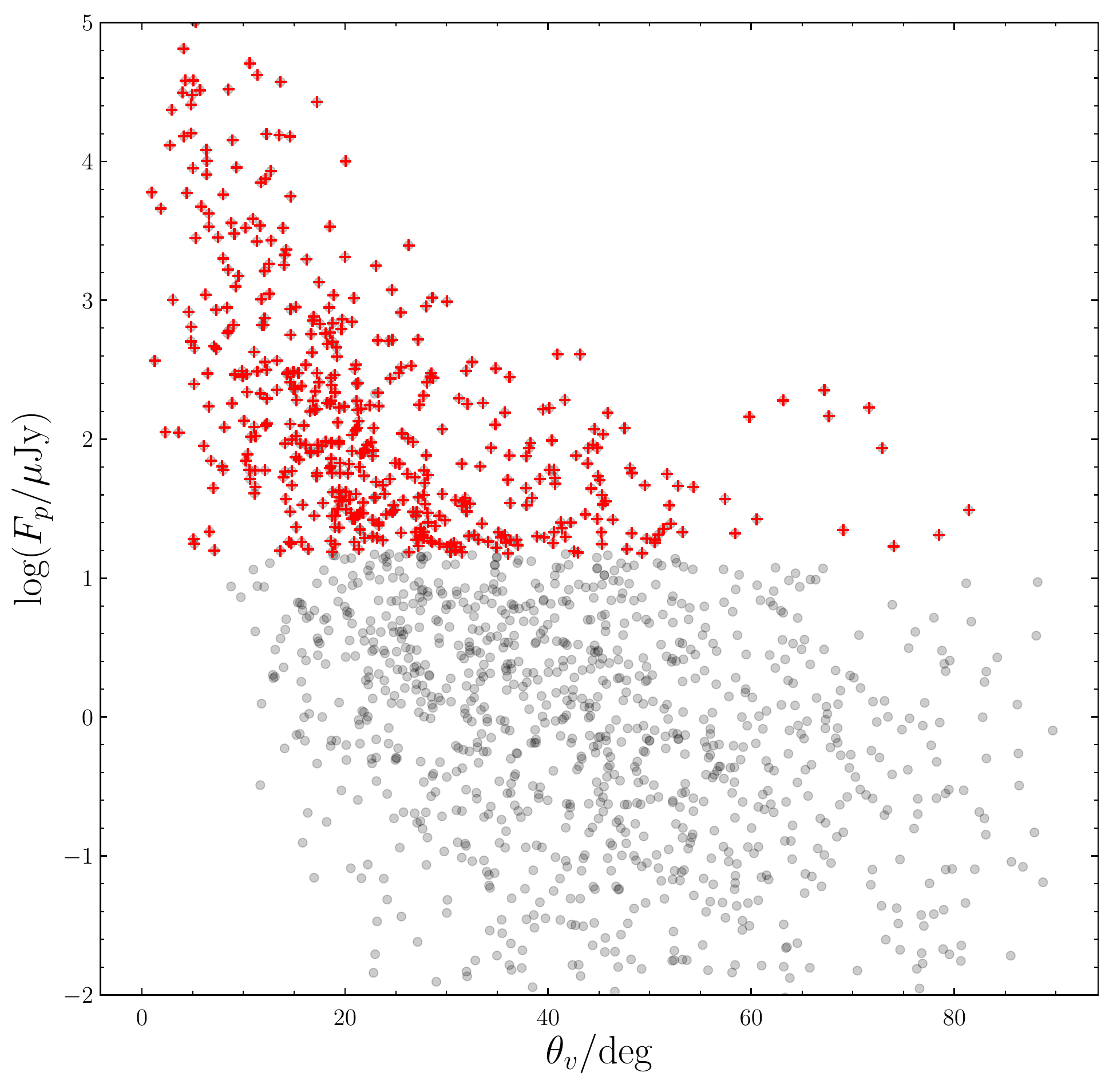}}
\resizebox{0.9\hsize}{!}{\includegraphics{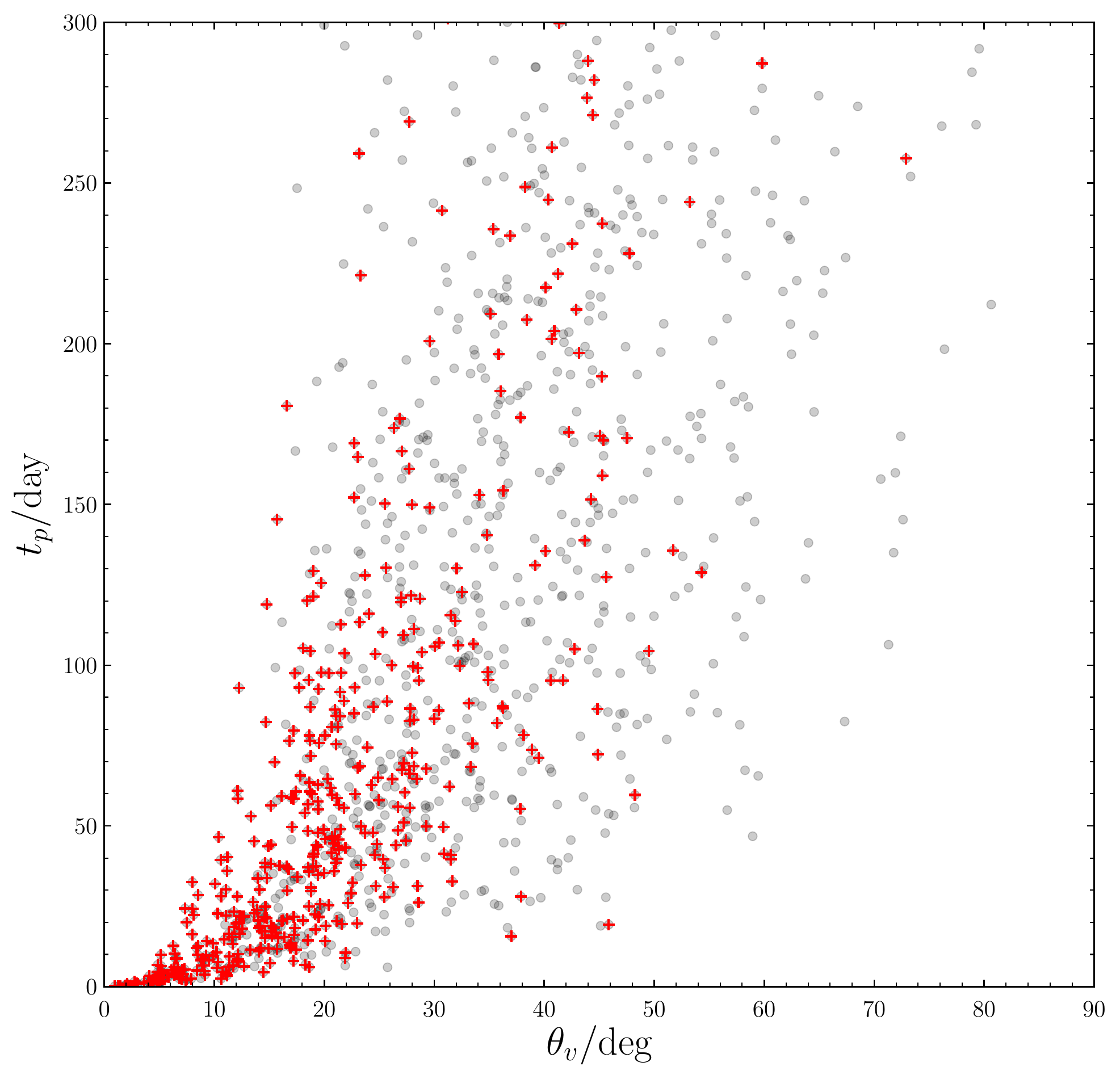}
\includegraphics{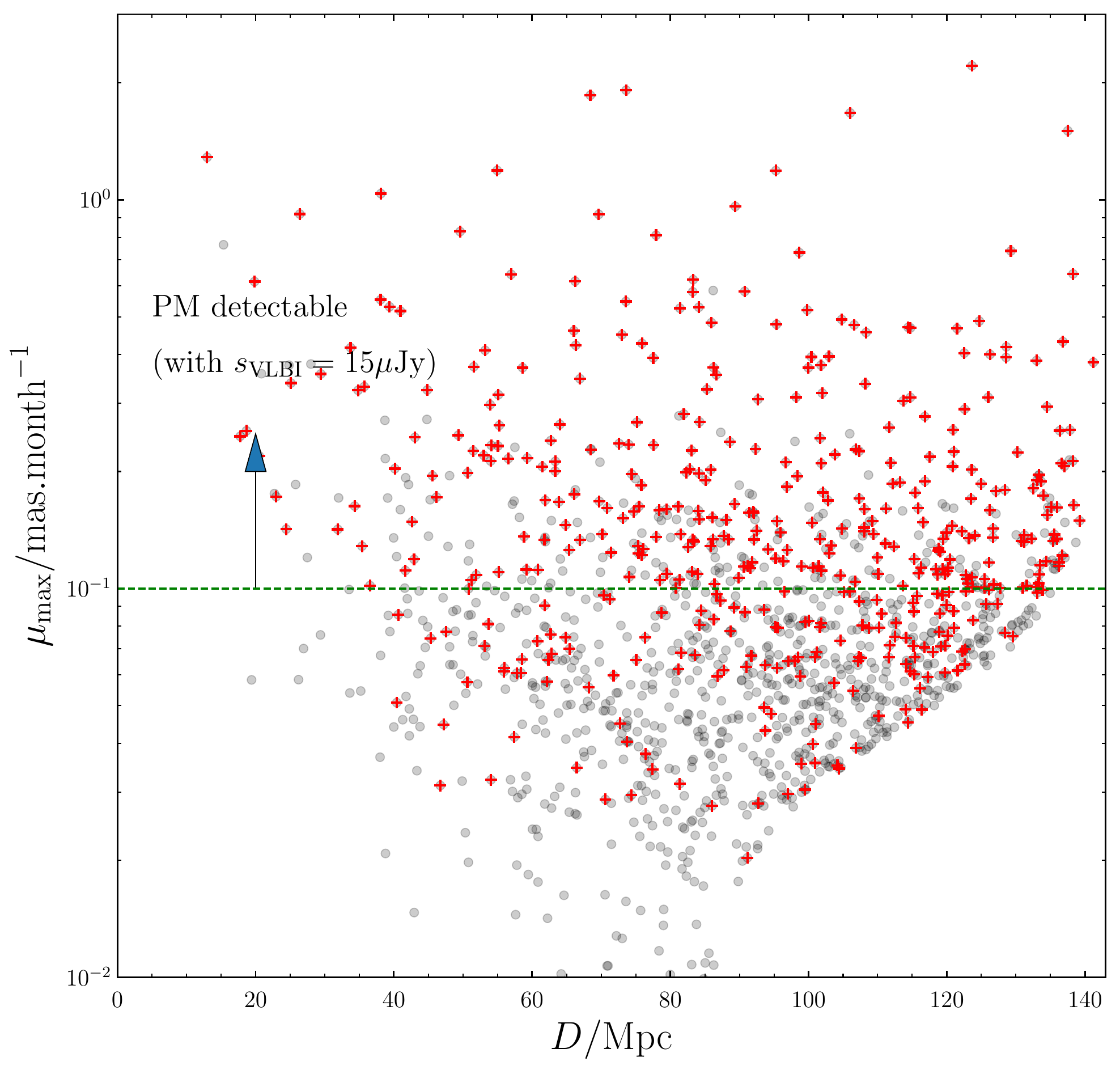}}
\end{center}
\caption{Predicted samples of GW-detected (dots) and jointly detected (crosses) mergers in the planes of various pairs of observables, for the O3+VLA configuration.
\textit{Upper left}: Distance and viewing angle, as well as the maximum distance to which the proper motion can be measured ($D_{\rm max}(\theta_{\rm v})$, see Eq.~\ref{eq:dmax}, green dotted line). \textit{Upper right}: Peak flux and viewing angle. \textit{Lower left}: Time of radio afterglow peak (assuming lateral expansion of the jet) and viewing angle. \textit{Lower right}: Maximum remnant proper motion and distance, as well as the limiting proper motion $\mu_{\rm lim}$ (green dotted line). }
\label{fig:2d}
\end{figure*}

We finally represent in Fig.~\ref{fig:2d} various plots connecting two observable quantities: $(\theta_{\rm v},D)$, $(\theta_{\rm v},F_{p, 3\,{\rm GHz}})$, $(\theta_{\rm v},t_p)$ and $(D,\mu_{\rm max})$. The grey dots correspond to events detectable in GW and the red \new{crosses} to those detectable in both GW and radio. In the $(\theta_{\rm v},D)$ diagram, the limiting distance for the measurement of the proper motion 
(Eq.~\ref{eq:dmax}) is shown. This figure illustrates the various observational biases discussed in this section, and especially the bias towards small viewing angles, mainly due to the limiting sensitivity of radio telescopes.

\subsubsection{Kinetic energy, external density and microphysics parameters}
\begin{figure*}
\resizebox{\hsize}{!}{\includegraphics{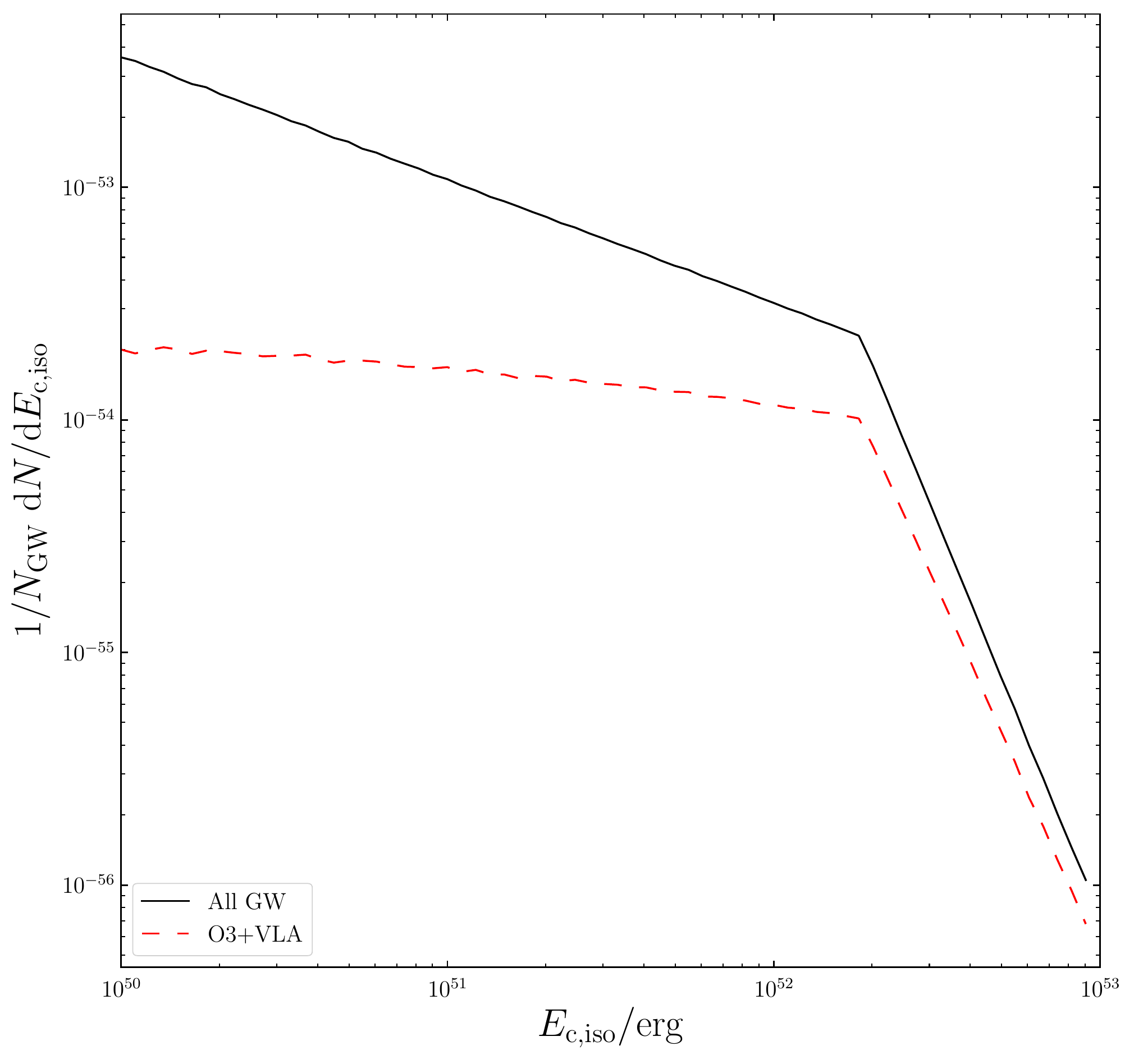}\includegraphics{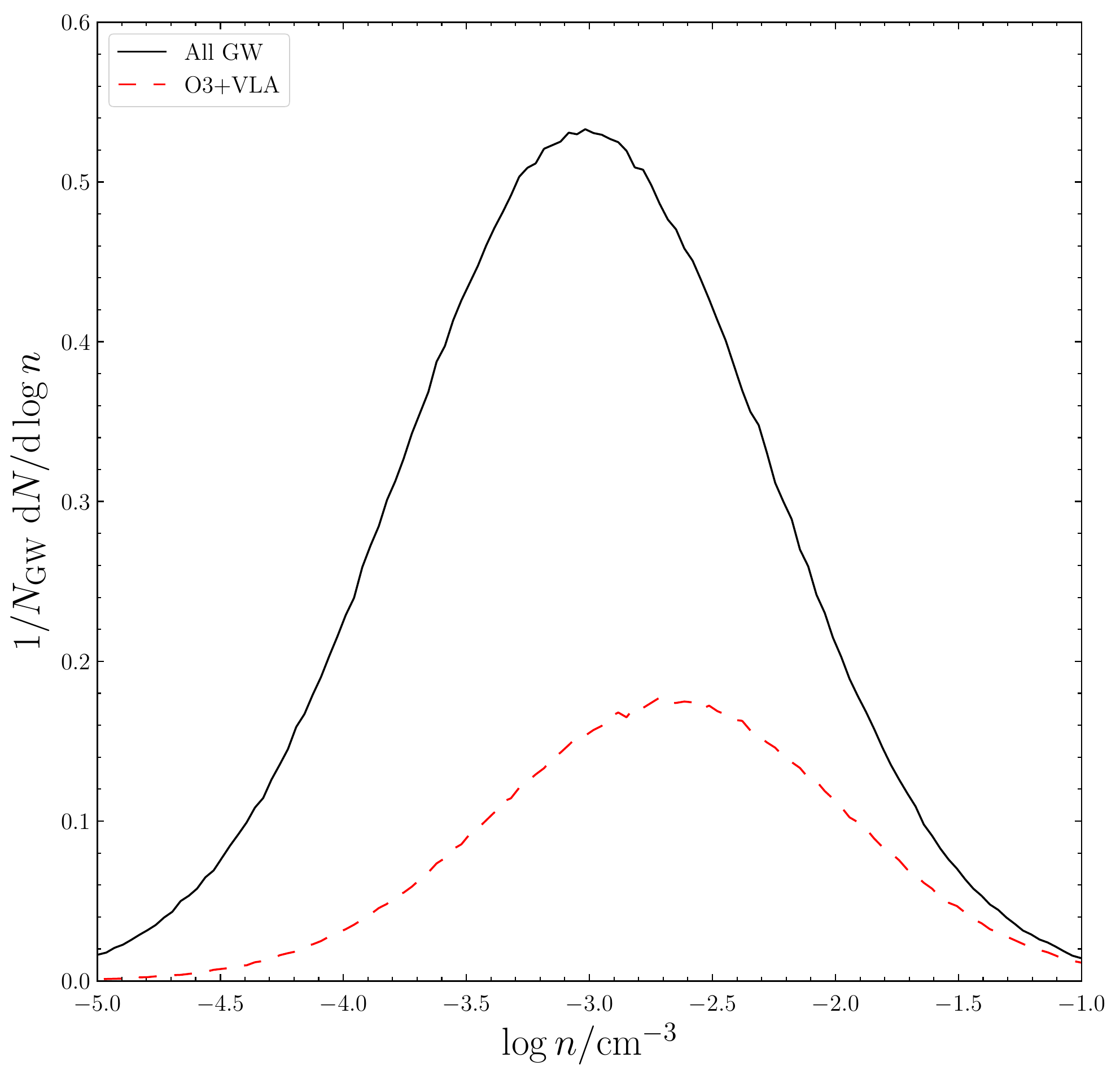}}
\caption{Differential distribution of the core jet kinetic energy (left) and external medium density (right) of the entire GW population (full) and the jointly detected events (dashed), in the O3+VLA  configuration. \new{Distributions are normalized to the fraction of jointly detected events among all GW-detected events, see text for details.}}
\label{fig:En}
\end{figure*}

The distributions of 
the 
kinetic energy $E_\mathrm{iso,c}$
and the external density $n$,
which are not directly observable but may result from fits of the afterglow, are shown if Fig.~\ref{fig:En}. 
In each case, the distributions 
of jointly detected events are compared
to the total GW detections. As the GW selection is independent of the kinetic energy and density, these distributions are the intrinsic distributions of our population model (Tab.~\ref{tab:fiducial}). 
As could be expected, the detection of mergers leading to a relativistic core jet with 
a large $E_\mathrm{iso,c}$ is favored. In jointly \new{detectable} mergers, the energy distribution approximately remains a broken power-law but the respective indices below and
above the break decrease from 
$0.53$ to $0.2$ and $3.4$ to $3.2$
respectively. \new{After applying the joint detectability criterion}, the external density distribution is simply shifted to higher densities, as the mean value is increased by a factor of $\sim 2.2$. The distribution of the microphysics parameter $\epsilon_\mathrm{B}$ is very similar to that of the density, because both parameters appear with the same power in the peak flux (Eq.~\ref{eq:FP}).

We find that the Lorentz factor of the jet at the time of peak flux has a median value of $\sim 4$. \new{This is consistent with the approximate jet core de-beaming relation $\Gamma \sim \left(\theta_\mathrm{v}-\theta_\mathrm{j}\right)^{-1}$ for a jet with $\theta_\mathrm{j}=0.1~{\rm rad}$ seen with the mean viewing angle $\theta_\mathrm{v}\sim 24^\circ$ found for the O3+VLA configuration}. This justifies to keep microphysics parameters representative of shock-acceleration in the ultra-relativistic regime for the whole population.

\begin{figure*}
\resizebox{\hsize}{!}{\includegraphics{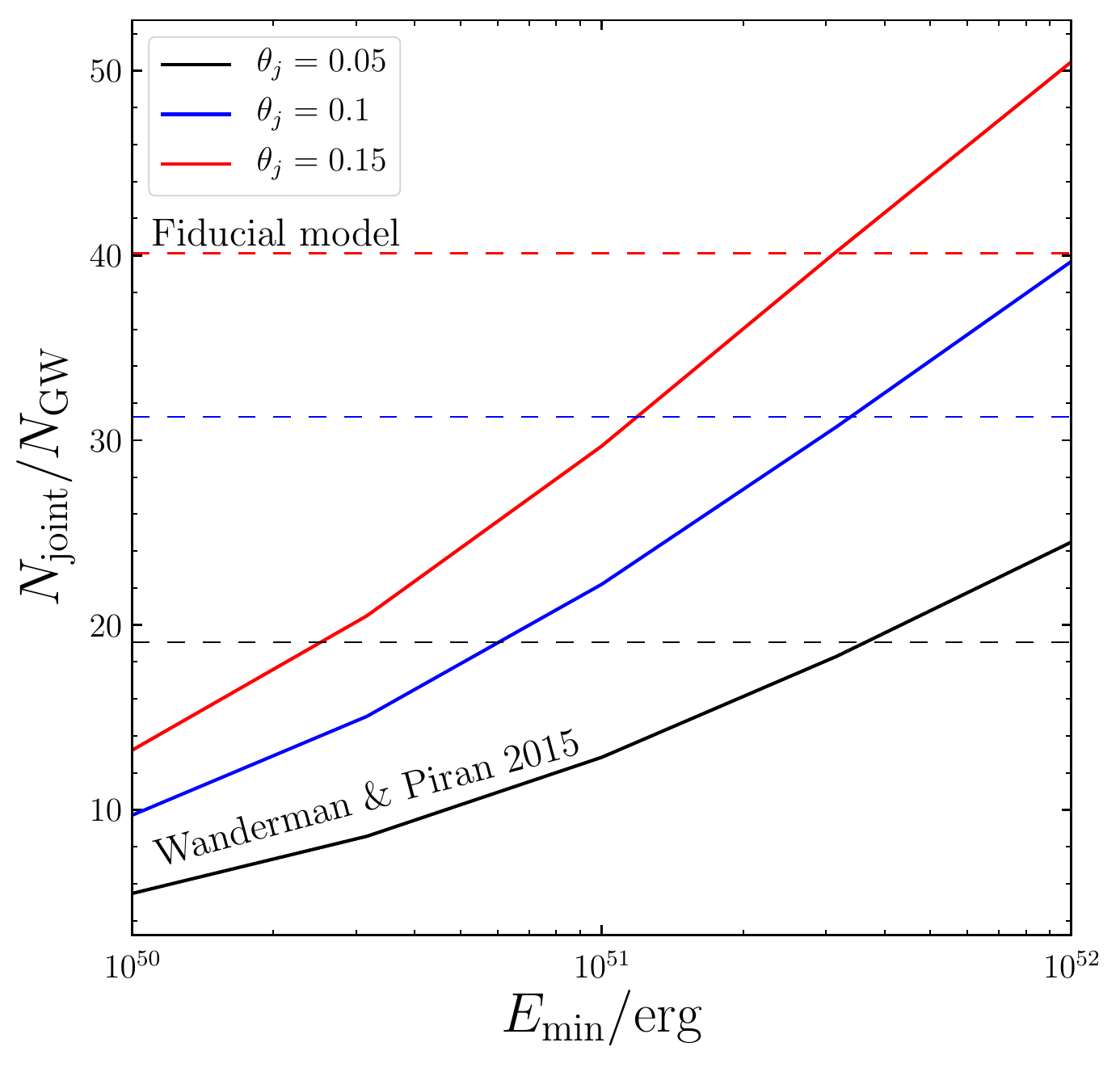}\includegraphics{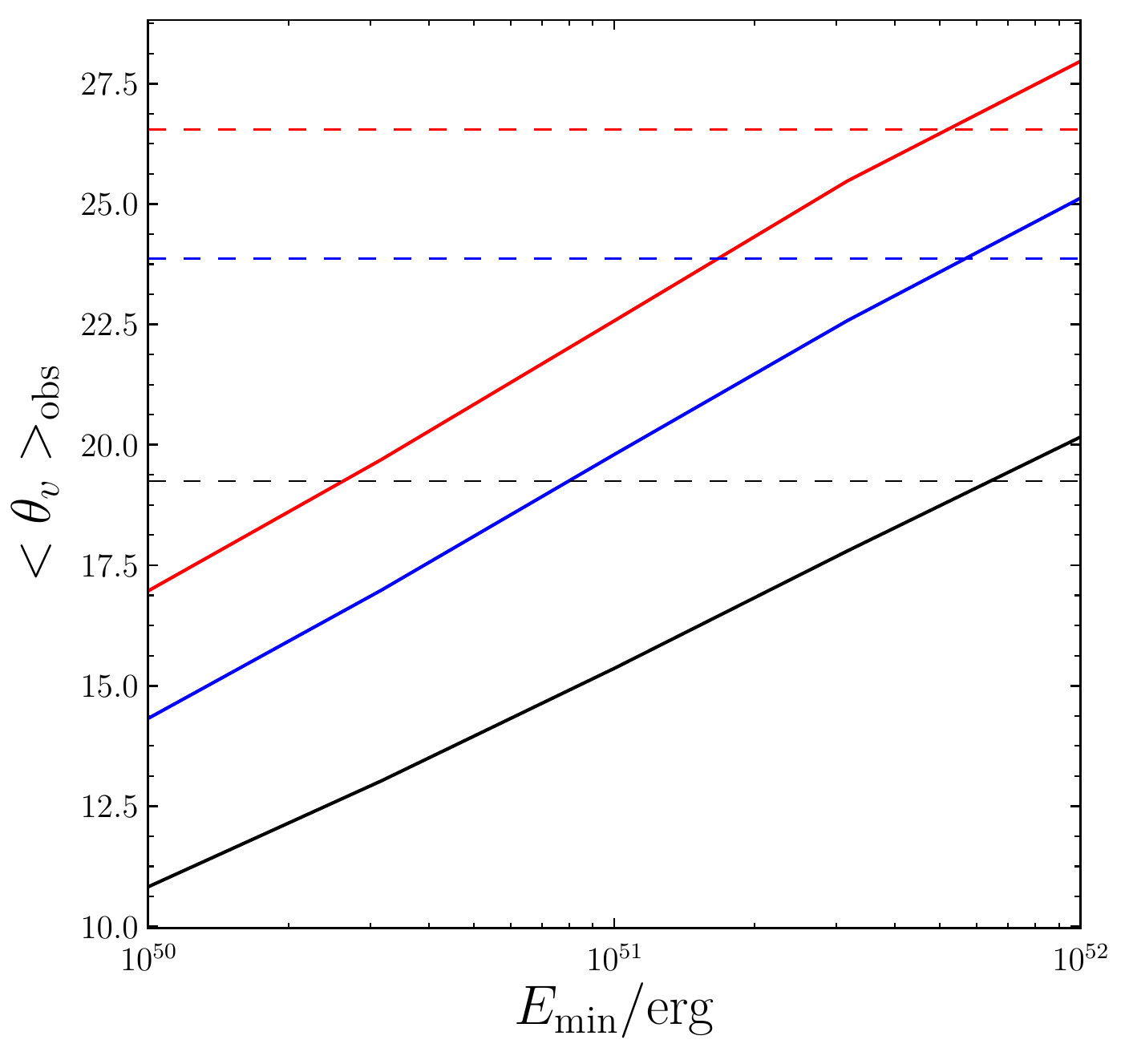}
\includegraphics{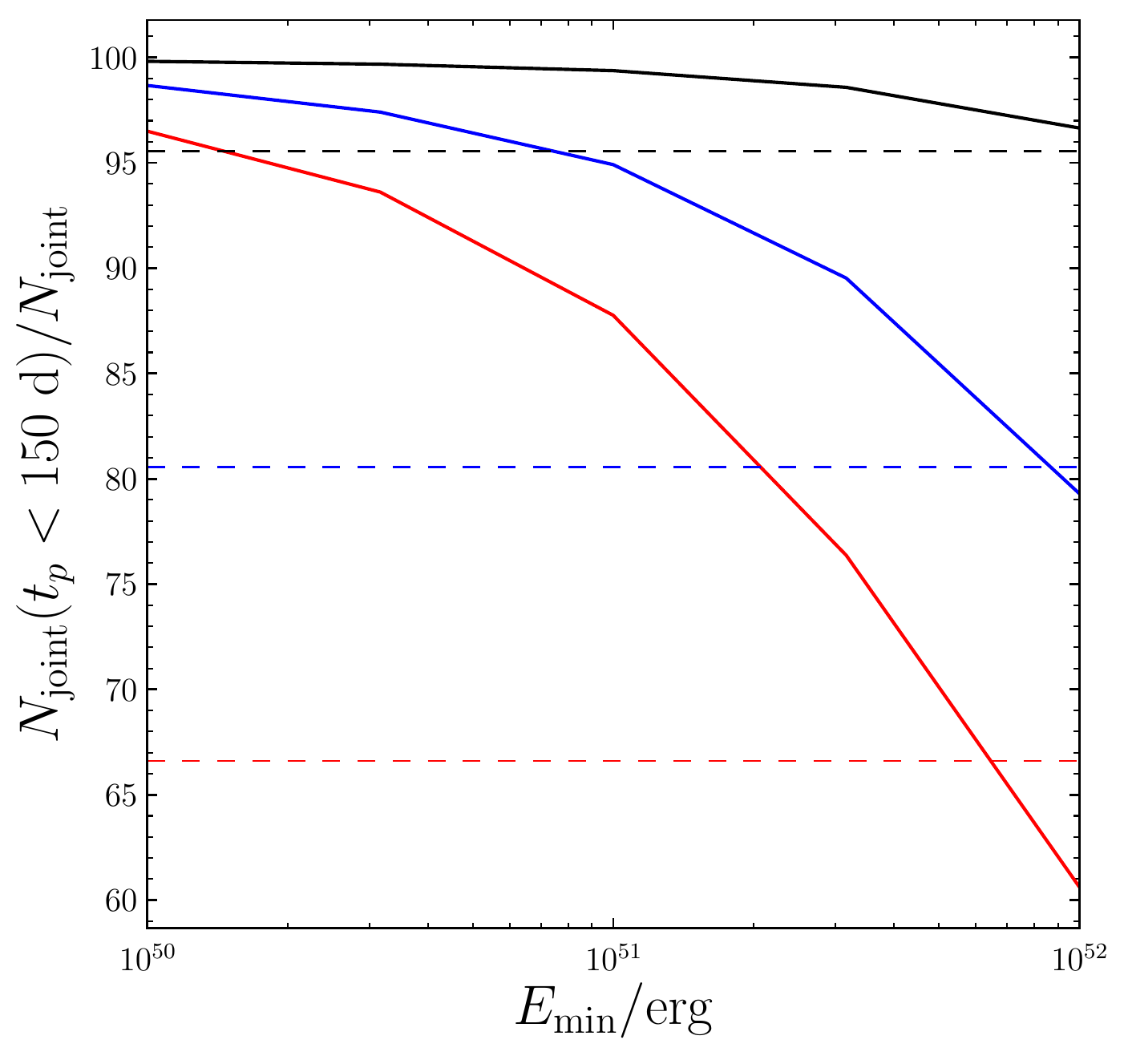}}
\caption{\textit{Left}: Fraction of joint events among GW events under the assumption of the short GRB luminosity function from \cite{105} for different values of the minimum kinetic energy $E_m$ (full lines), and for our fiducial model (dashed lines), in the O3+VLA configuration. The results are plotted for different values of the jet opening angle $\theta_{\rm j}=0.05$, $0.1$ and $0.15$ rad (red, black and blue lines respectively). \textit{Middle}: Mean viewing angle (same color coding). \textit{Right}: Fraction of joint events with peaks earlier than 150 days post-merger, with the hypothesis of jet lateral expansion, same color coding.}
\label{fig:fig9}
\end{figure*}

\subsection{Impact of the population model uncertainties on the predicted population}
\label{sec:variations}
\subsubsection{Jet energy distribution and opening angle}
We now consider alternatives to our fiducial population model.
With a distribution of jet energy that increases below the break
(i.e. taking $\alpha_1<0$ in Eq.~\ref{eq:phiE}) as would be obtained from the short GRB
luminosity function of \cite{105} ($\alpha_1=-0.9$) and still assuming the same standard values for the duration and efficiency, the fraction of events \new{detectable with the VLA} would typically be three times smaller: 15\% for O2 and 10\% for O3 compared to 43\% and 31\% with $\alpha_1=0.5$. The viewing angle and peak times would also be strongly affected, showing \new{mean} $\theta_v$'s for O2 (resp. O3) at only $17^\circ$  
(resp. $14^\circ$), 98\% (resp. 99\%) of afterglows peaking before
150 days \new{in the jet expansion hypothesis, and 90\% (resp. 93\%) without lateral expansion}.

These results were obtained assuming a typical value $f_\gamma=20\%$ for the gamma-ray efficiency but one cannot exclude that $f_\gamma$ may vary from
from less than 1\% to more than 20\%. Estimates of $f_\gamma$ from 
afterglow fitting have been indeed found to cover a large interval \citep[e.g.][]{47, 48}. 
Thus, the three orders of magnitude in isotropic gamma-ray luminosity could partially 
result from differences in efficiency, the jet isotropic kinetic energy being
restricted to a smaller interval. Fig.~\ref{fig:fig9} shows the effect of increasing the 
minimum kinetic energy $E_\mathrm{min}$ from $10^{50}$ to $10^{52}$~erg in the energy function deduced from the short GRB luminosity function of \cite{105} on the detected fraction,
the mean viewing angle and the fraction of afterglows to peak
before 150 days. As could be expected, when the minimum kinetic energy is increased, the results become closer to those of the fiducial model where the energy distribution function decreases below the peak.  

We also show in Fig.~\ref{fig:fig9} the effect of changing the typical value of the jet opening
angle from 0.1 to 0.05 or 0.15~rad. When $\theta_{\rm j}=0.05$ (resp. $0.15$), the detected fraction decreases (resp. increases)
to 19\% (resp. 40\%), the mean viewing angle decreases (resp. increases) by $4^\circ$, and the peak of the light curve generally occurs later (resp. earlier).

\subsubsection{External density distribution and microphysics}

\new{The afterglow peak flux depends greatly on the density of the circum-merger medium}, as can be seen from Eq.~\ref{eq:FP}. \new{Therefore, the ability to detect the afterglows of a population of mergers thus depends strongly on the density of the media hosting the binaries upon merger.} This is illustrated in Tab.~\ref{tab:n0}, where we give this fraction for different external density distribution central values. \new{These figures were calculated using the semi-analytical model (Eq.~\ref{eq:FluxApp}), taking into account the full synchrotron spectrum including self-absorption.}

If there is a \new{significant population of fast-merging neutron star binaries}, as proposed by e.g. \cite{157, 156, 158, 155}, these should merge in higher density environments, producing brighter afterglows and can notably contribute to the observed population
even if there are intrinsically less numerous. With a density distribution centered at $n=10$ cm$^{-3}$
and the other parameters taking their fiducial values, 99\% of the GW events produce detectable radio afterglows \new{in the O3+VLA configuration}. 

The microphysics parameter $\epsilon_{\rm B}$ enters in Eq.~\ref{eq:FP} with the same power as the external density.
Moreover our fiducial choices for the distribution of these two quantities are nearly the same (with 
simply a limit in range for $\epsilon_{\rm B}$) so that changing its central value  
affects the detected fraction in the same way. 


\section{Discussion and conclusion}
\label{sec:discussion}

We studied the population of binary neutron star mergers to be observed jointly through GW and their radio afterglows in future multi-messenger observing campaigns. For this we have assumed a likely population of mergers inspired by prior short GRB science, and simulated the GW and jet-dominated radio afterglow emissions from these. We have made predictions on the rates of such future events, and on how the observables from these events are expected to distribute themselves.

In the case of the ongoing O3 run of the LVC and assuming our fiducial set of parameters (see Tab.~\ref{tab:fiducial}), we predict that $\sim 30\%$ of GW events should have a radio afterglow \new{detectable by the VLA}. These joint events should have mean viewing angles of $\sim 24^\circ$, and should peak earlier than 150~days post-merger in $\sim 80\%$ of \new{cases assuming lateral expansion of the jet}.

As illustrated in \new{the 170817} event, the lateral material may also contribute to the peak flux of the radio afterglow. Fig.~\ref{fig:radiofit} (right) shows that this may increase the peak flux by a factor of up to 1.5. In this case, our O3 predictions are revised to a fraction of 36\% of events with detectable afterglow and a similar mean viewing angle of $25^\circ$.

\begin{table}[!t]
\caption{Fraction of joint events among GW events for different intrinsic density distribution central values, for the O3+VLA detector configuration, as calculated with the semi-analytical model (Eq.~\ref{eq:FluxApp}).}
\begin{center}
\begin{tabular}{ll}
\hline
\hline
$n_0/{\rm cm}^{-3}$ & $N_{\rm joint} / N_{\rm GW}$ \\
\hline
$10^{-4}$ & 22 \% \\
$10^{-3}$ & 40 \%  \\
$10^{-2}$ & 63 \% \\
$0.1$ & 83 \% \\
$1$  & 94 \% \\
$10$  & 97 \% \\
\hline
\end{tabular}
\end{center}
\label{tab:n0}
\end{table}

\subsection{Detecting the detectable events}
\new{Throughout this work}, we have applied a threshold on the GW SNR and radio afterglow peak flux to label an event as jointly detected. Thus our study concerns the class of \textit{detectable} events. The fraction of these detectable events that can \new{actually} be detected depends strongly on the accuracy of the localization, which is related to the size of the localization skymap provided by the GW network, the possible detection of an associated GRB and the efficiency of the search for a kilonova, in terms of covered sky area and limiting magnitude.

\new{
Assuming that all neutron star mergers produce a kilonova and even taking into account a probable anisotropy of the emission, it is expected that 100\% of kilonovae are \textit{detectable} during O3 for a limiting magnitude of 21 \citep{groningen, inprep}.   
However, the search for the kilonova is complicated due to the large localization skymaps of GW events. These error boxes can be improved in case of the detection of an associated GRB, which however should remain rare (see Sec.~\ref{sec:4.1.2} and \citealt{67}). During O3,  existing facilities with large fields of view in the optical are experiencing difficulties covering 100\% of $ \gtrsim 1000\, \mathrm{deg}^2 $ skymaps with a deep limiting 
magnitude, especially in the southern sky. Instruments such as the the Zwicky Transient Facility \citep{ZTF} cover the northern sky down to magnitude $r\sim~20.5$.}

\new{On the other hand, when the LIGO/Virgo instruments will have reached their design sensitivities, it is expected that one or two new interferometers will have joined the GW network: KAGRA \citep{KAGRA} and Ligo-India \citep{LINDIA}. This should lead to a significant improvement of the localization. In addition, the deep coverage of even large error boxes will be facilitated by new facilities such as the Large Synoptic Survey Telescope \citep{LSST}.
Nevertheless, there are additional difficulties in the search for the kilonova: kilonova/host galaxy contrast at large distances, possible large offsets, availability of a photometric and spectroscopic follow-up of the candidates, etc. In this context, it is clear that the efficiency of the search for kilonovae will always remain below 100\%, but should increase with time and may potentially be high in the design configuration of the GW network. 
}

\new{Even assuming the} 
localization is acquired \new{thanks to the detection of a kilonova}, a continuous monitoring of the remnant up to $\sim 150~{\rm days}$ may be necessary to detect the peak of the \new{radio} afterglow, in the case of marginally detectable events. Even then, only events with peaks somewhat larger than the radio threshold should yield observations of astrophysical interest because extended observations of the rise, peak and decay of the afterglow are necessary to resolve the structure and dynamics (e.g. expansion) of the ultra-relativistic jet, as illustrated by the case of GRB170817A.

\new{As also illustrated in the case of GRB170817A}, VLBI measurements \new{are} instrumental to \new{assess the presence and} resolve the structure of the jet. Thus an even more restrictive criterion for full event characterization could be the detectability of the remnant proper motion. We have shown in Sec.~\ref{sec:4.1.4} that this would largely decrease the fraction of events, especially in the context of design GW detectors.

\subsection{Constraining the BNS population and short GRBs with radio afterglows}
We have studied the sensitivity of the expected observables' distributions to the population parameters as well as the radio and GW detector configurations. In particular, we have shown the uncertainties on the rates and typical viewing angles of these events inherited from the uncertainties on the luminosity function of short GRBs and of the density of the media hosting the merger.

In Sec.~\ref{sec:rates}~and~\ref{sec:4.1.2} we have discussed the combined influences of increasing the radio and GW detector sensitivity on the rate and viewing angle of the events. At this stage, both improvements would be beneficial as the predicted rate in the current O3+VLA \new{configuration} is well below 50\% of joint events among GW events.

As the GW horizon increases, more events will be seen on-axis. In the case of an opening angle of 0.1~rad and in the likely short-term evolution of the detector configuration from O3+VLA (current) to design+VLA (early 2020s), we predict that 5\% to 10\% of events should be seen on-axis, increasing the probability of a GRB counterpart in addition to the GW and afterglow. This figure is fully consistent with that announced in \cite{67} for events with a GW+GRB association. In fact, as a GW+classical GRB association is equivalent to an on-axis event, one may measure the opening angle of typical GRB jets by considering the ratio of GW events with afterglow counterpart only to events with both afterglow and GRB, thus exploiting the new possibility of observing afterglows without detecting the GRB.

More generally, a population of observed events corresponds to an intrinsic population of mergers. Thus a comparison of a statistical number of joint event observations to our predicted distributions is a means of measuring fundamental parameters of the population of mergers, and of constraining GRB quantities such as their energy function.

Finally, future facilities such as the SKA may bring detections of orphan radio afterglows through deep radio surveys. These would not be subject to the GW criterion and may probe another sub-population of mergers, bringing yet another class of constraints on these phenomena.

In conclusion, regardless of the evolution of GW and radio detectors, multi-wavelength afterglows will remain instrumental in the study of binary neutron star mergers as \new{windows} on both their environment and their role as progenitors of short GRBs. They will bring precious insight at the level of the population of mergers and on an event-to-event basis, provided the sources may be accurately localized in the sky.

\section*{Acknowledgments}
The  authors  acknowledge financial support from the Centre National d'\'Etudes Spatiales (CNES), and thank Paz Beniamini for useful discussions.

\bibliographystyle{aa} 
\bibliography{pop_paper}
\end{document}